\documentclass[10pt,conference]{IEEEtran}
\usepackage{cite}
\usepackage{amsmath,amssymb,amsfonts}
\usepackage{algorithmic}
\usepackage{graphicx}
\usepackage{textcomp}
\usepackage{xcolor}
\usepackage[hyphens]{url}
\usepackage{fancyhdr}
\usepackage{hyperref}

\usepackage{threeparttable}
\usepackage[ruled,vlined]{algorithm2e}
\usepackage{array}
\usepackage{makecell}
\usepackage{tabularx,colortbl}
\usepackage{booktabs}

\title{3D-MoE: Complementarity-Guided Hardware–Software Co-Design for MoE Decode on 3D-Stacked NMP}

\author{
  \IEEEauthorblockN{%
    Yudong Pan\IEEEauthorrefmark{1}\IEEEauthorrefmark{2},
    Lian Liu\IEEEauthorrefmark{1}\IEEEauthorrefmark{2},
    Shixin Zhao\IEEEauthorrefmark{1}\IEEEauthorrefmark{2},
    Yintao He\IEEEauthorrefmark{1}\IEEEauthorrefmark{2},
    Kaiyan Chang\IEEEauthorrefmark{1}\IEEEauthorrefmark{2},
    Tianhua Han\IEEEauthorrefmark{1}\IEEEauthorrefmark{2},\\
    Chenyang Ai\IEEEauthorrefmark{3},
    Haonan Zhu\IEEEauthorrefmark{1}\IEEEauthorrefmark{2},
    Mengdi Wang\IEEEauthorrefmark{1}\IEEEauthorrefmark{2},
    Yinhe Han\IEEEauthorrefmark{1}\IEEEauthorrefmark{2} and
    Ying Wang\IEEEauthorrefmark{1}\IEEEauthorrefmark{2}\IEEEauthorrefmark{4}
  }
  \IEEEauthorblockA{%
    \IEEEauthorrefmark{1}SKLP, Institute of Computing Technology, Chinese Academy of Sciences, Beijing, China\\
    \IEEEauthorrefmark{2}University of Chinese Academy of Sciences, Beijing, China\\
    \IEEEauthorrefmark{3}University of Edinburgh, Edinburgh, United Kingdom\\
    \IEEEauthorrefmark{4}Corresponding author: wangying2009@ict.ac.cn
  }
  \IEEEauthorblockA{%
    Email: \{panyudong23,liulian211,zhaoshixin18,hantianhua22,zhuhaonan24\}@mails.ucas.ac.cn,\\
    \{heyintao19z,wangmengdi,yinhes,wangying2009\}@ict.ac.cn,
    changkaiyan@live.com, C.Ai-4@sms.ed.ac.uk
  }
}

\begin{document}
\maketitle
\pagestyle{plain}
\thispagestyle{plain}

%%%%%%%%%%%%%%%%%%%%%%%%%%%%%%%%%%%%%%%%
%%%%%%%% -- PAPER CONTENT STARTS -- %%%%%%%%%

\begin{abstract}

Mixture-of-Experts (MoE) have become a mainstream way to scale Large Language Models (LLMs), because each token activates only a small subset of experts, reducing inference cost. However, MoE decode remains highly memory-bound and exhibits strongly skewed, time-varying expert loads. 3D-stacked DRAM near-memory processing (NMP) provides the bandwidth that decode demands, yet its tiled organization makes load balancing expensive: expert parallelism (EP) preserves locality but suffers from hotspot-induced imbalance, while more distributed strategies weaken locality and increase NoC traffic.

To address this locality-balance conflict, we observe that domain-specialized token routing creates offsetting expert-load fluctuations across decode steps, a property we call expert complementarity. Using this insight, we propose 3D-MoE, a hardware--software co-designed system for MoE decode on 3D-stacked NMP. 3D-MoE uses offline, one-time complementarity-guided grouping and locality-aware mapping to establish coarse-grained inter-group balance, handles residual intra-group hotspots through a lightweight Collaborative Memory-Network (CMN) that streams each hot expert's weight tiles once from its host DRAM channel to helper cores over a reconfigured local broadcast path, and uses cost-aware online remapping to remain robust to workload drift. Across MoE models, 3D-MoE achieves an average 4.63$\times$ MoE-layer speedup over EP and improves end-to-end throughput by 1.58$\times$--1.86$\times$ over the strongest NMP baselines, with 2.11\% additional logic-die area.
\end{abstract}

\section{Introduction}

Mixture-of-Experts (MoE) architectures have become mainstream for scaling Large Language Models (LLMs): each token activates only a small expert subset, decoupling model capacity from inference cost~\cite{liu2024deepseek3,jiang2024mixtral}. Yet this sparsity makes autoregressive decoding highly memory-bound; each expert receives few tokens per step, producing small GEMMs with poor data reuse~\cite{zhong2024distserve}. In our profiling, MoE layers account for 83--92\% of end-to-end latency at batch size 64 and dominate serving latency~\cite{huang2023towards}. Expert loads are also highly skewed and dynamic, with the hottest expert receiving over 20$\times$ the step-average token count~\cite{fang2025klotski}.

\begin{sloppypar}
3D-stacked DRAM near-memory processing (NMP) is well matched to bandwidth-dominated MoE decode by serving expert weights directly from local stacked DRAM~\cite{yun2024duplex,gao2017tetris,tian2023abndp}. Yet its tiled organization creates a severe cost asymmetry: local accesses are fast, whereas remote accesses traverse the network-on-chip (NoC) and consume shared bandwidth. MoE deployment consequently faces a three-way tension among locality, load balance, and communication cost. Expert Parallelism (EP) best matches this structure by keeping expert weights local, but suffers from routing skew. Tensor Parallelism (TP) smooths load, yet turns every expert invocation into a distributed operation with synchronization and NoC overhead. Dynamic EP methods partially alleviate hotspots, but the difficulty of predicting expert loads and handling remote hotspots limits their benefits on NMP~\cite{deepseek_eplb_2025,fang2025klotski}. \textit{The central challenge is therefore to retain EP's locality while mitigating its imbalance.}
\end{sloppypar}

Our key observation is that MoE routing skew is not purely random. Although MoE training often includes load-balancing objectives, pre-trained models develop structured expert preferences, often reflected in domain-skewed activation patterns~\cite{pan2025stratum,yang2025qwen3}. As the request mix changes, demand shifts across expert subsets and induces offsetting variations in their decode-step loads. We call this trace-level load cancellation \textbf{expert complementarity}. This structural signal suggests an EP-friendly balancing mechanism: grouping complementary experts and mapping them to nearby NMP cores can stabilize aggregate group load without sacrificing weight locality.

Turning complementarity into a practical deployment primitive remains non-trivial. First, highly volatile step-level token routing obscures the useful complementarity signal, requiring a robust method to distinguish persistent relations from transient noise. Second, complementarity improves balance mainly at the \textit{group level}, leaving transient hotspots within a group. Because a hot expert's weights remain pinned to its host core, nearby helper cores require repeated remote weight fetches or inefficient packet-routed one-to-many delivery. Third, the offline expert deployment derived from historical traces must remain robust under workload drift, calling for lightweight online correction rather than frequent global re-optimization.

\begin{sloppypar}
To address these challenges, we propose 3D-MoE, a hardware--software co-design for MoE decode on 3D-stacked NMP. 3D-MoE exploits complementarity hierarchically. First, it profiles decode-step expert loads on representative mixed workloads to filter transient routing noise and expose persistent complementary structure in the traces. It then performs one-time \textbf{complementarity-guided expert grouping} and mapping through a two-stage offline optimization that accounts for inter-group balance, expected residual collaboration cost, and communication-aware placement.
Second, we introduce the \textbf{Collaborative Memory-Network (CMN)}, a lightweight co-design of the memory controller and NoC that handles residual intra-group hotspots through a specialized ``load once, broadcast to many'' path: the host reads each hot expert's weight tile once from local DRAM and streams it to helper cores over a temporary broadcast path created by lightweight NoC reconfiguration. Third, tiered \textbf{cost-aware online remapping} applies frequent low-cost adjustments while reserving higher-cost reconfigurations for long-term workload drift.

\end{sloppypar}

This paper makes the following four contributions:
\begin{itemize}
\item We identify expert complementarity as a trace-level load-cancellation structure in MoE decode, and validate that it is sufficiently stable across datasets and time windows to support one-time offline deployment.

\item We formulate complementarity-guided expert grouping and locality-aware mapping as sequential optimization stages that balance inter-group load while accounting for intra-group collaboration and routing costs.
\item We design CMN, a lightweight hardware fast path for residual intra-group hotspot assistance that reduces redundant DRAM reads and per-hop packet processing.

\item We integrate offline optimization, localized collaborative execution, and tiered online correction into 3D-MoE, achieving 4.63$\times$ average MoE-layer speedup over EP and 1.58$\times$--1.86$\times$ end-to-end throughput gains over the strongest NMP baselines with only 2.11\% area overhead.
\end{itemize}

\section{Background and Design Implications}

\subsection{MoE Architecture \& Decode Behavior}
\label{subsec:2.1}
\begin{figure}[tp]
    \centering
    \includegraphics[width=0.88\linewidth]{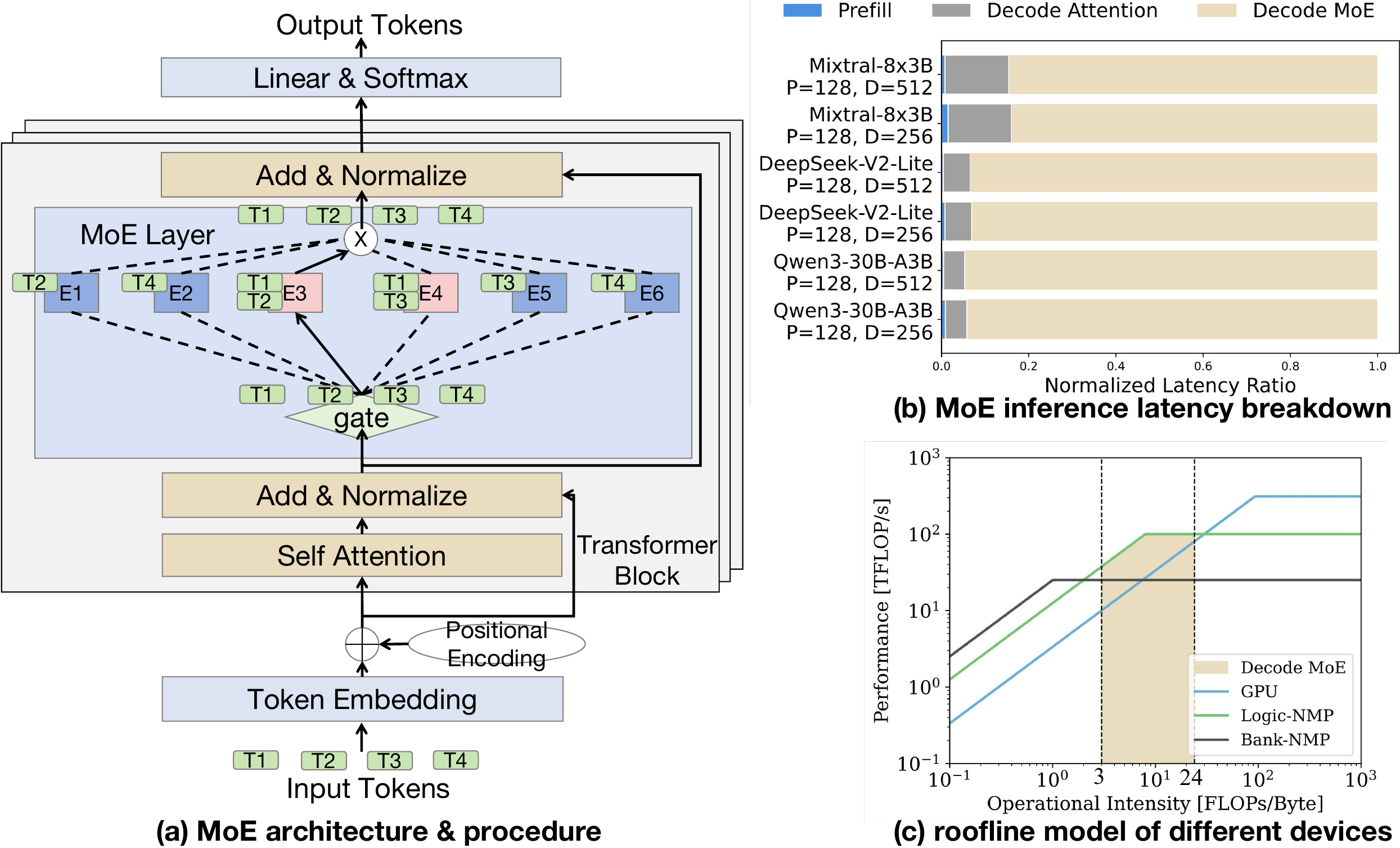}
    \caption{MoE workflow, latency breakdown, and device roofline comparison.}
    % \vspace{-0.7em}
    \label{fig:moe_structure}
\end{figure}

\begin{figure}[tp]
    \centering
    \includegraphics[width=0.9\linewidth]{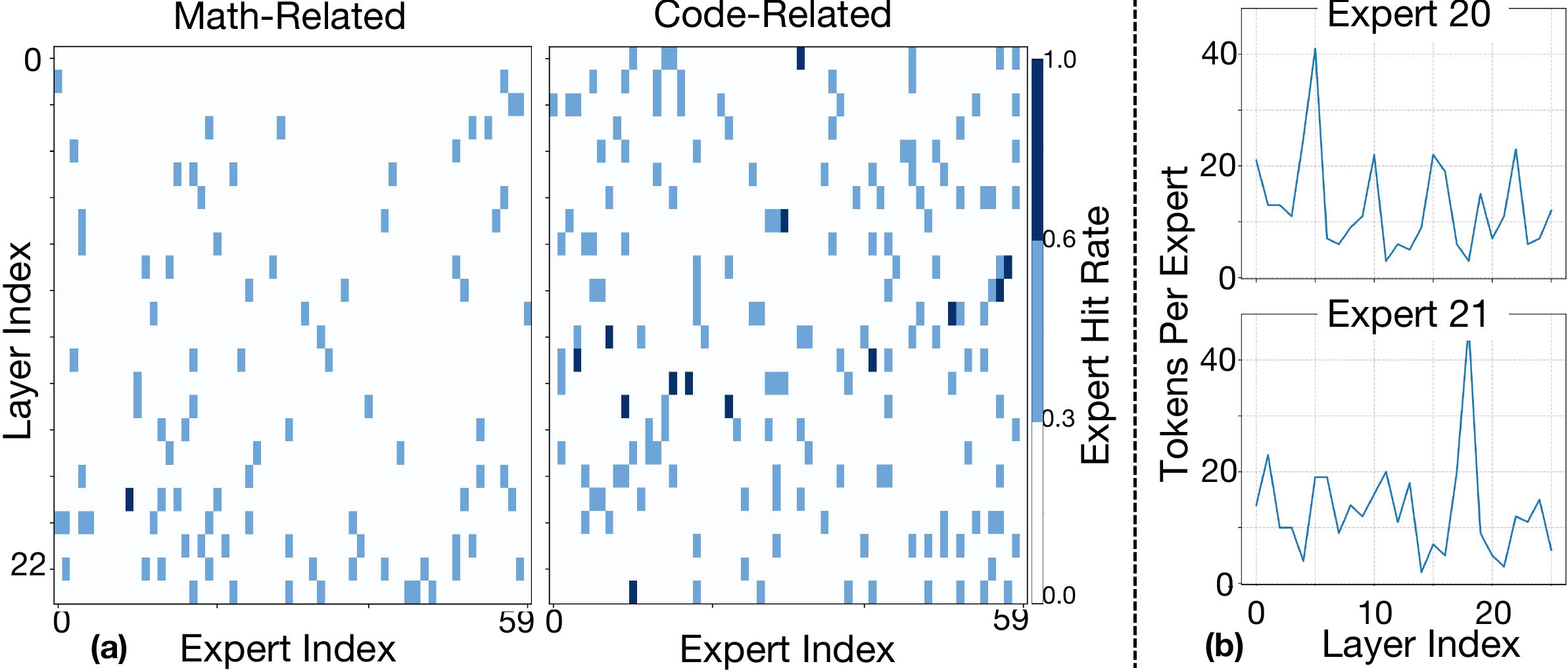}
    \caption{Expert activation analysis on Qwen3-30B-A3B. (a) Expert hit rates across domains. (b) Inter-layer variation.}
    \label{fig:expert_act_hot_diff}
\end{figure}

% TODO 这里引用可能需要改一下
Mixture-of-Experts (MoE) has become a pivotal architecture for scaling Large Language Models (LLMs) efficiently~\cite{liu2024deepseek2,liu2024deepseek3,jiang2024mixtral,yang2025qwen3,du2022glam}. As shown in Fig.~\ref{fig:moe_structure}(a), it replaces each Transformer Feed-Forward Network (FFN) with parallel experts. A lightweight gate selects only a few experts for each token, scaling model capacity while keeping per-token computation bounded~\cite{chiang2024chatbot,liu2024deepseek3}. Training-time load-balancing objectives do not eliminate expert specialization: pre-trained models often develop domain-preferred experts~\cite{li2023accelerating,yang2025qwen3}, so different request domains activate different expert subsets rather than all experts uniformly (Fig.~\ref{fig:expert_act_hot_diff}(a)).

Efficient MoE inference remains challenging, especially during autoregressive decode. Even under batching, each expert receives few tokens per step, fragmenting execution into small GEMMs or GEMVs with poor data reuse and low arithmetic intensity~\cite{huang2023towards}. MoE layers consequently account for $83$--$92\%$ of end-to-end latency at batch size 64 in our profiling, consistent with prior studies~\cite{tang2024hobbit}, as shown in Fig.~\ref{fig:moe_structure}(b). Meanwhile, expert loads are highly skewed and rapidly changing~\cite{fang2025klotski}. As shown in Fig.~\ref{fig:expert_act_hot_diff}, only a few experts become hot at a step; the hottest expert can receive over $20\times$ the step-average load, while each expert's own load varies by $2$--$7\times$ across steps and layers. MoE decode therefore poses two coupled challenges: it is highly memory-bound, yet its expert loads are also severely imbalanced and time-varying.

\begin{figure}[tp]
    \centering
    \includegraphics[width=1.0\linewidth]{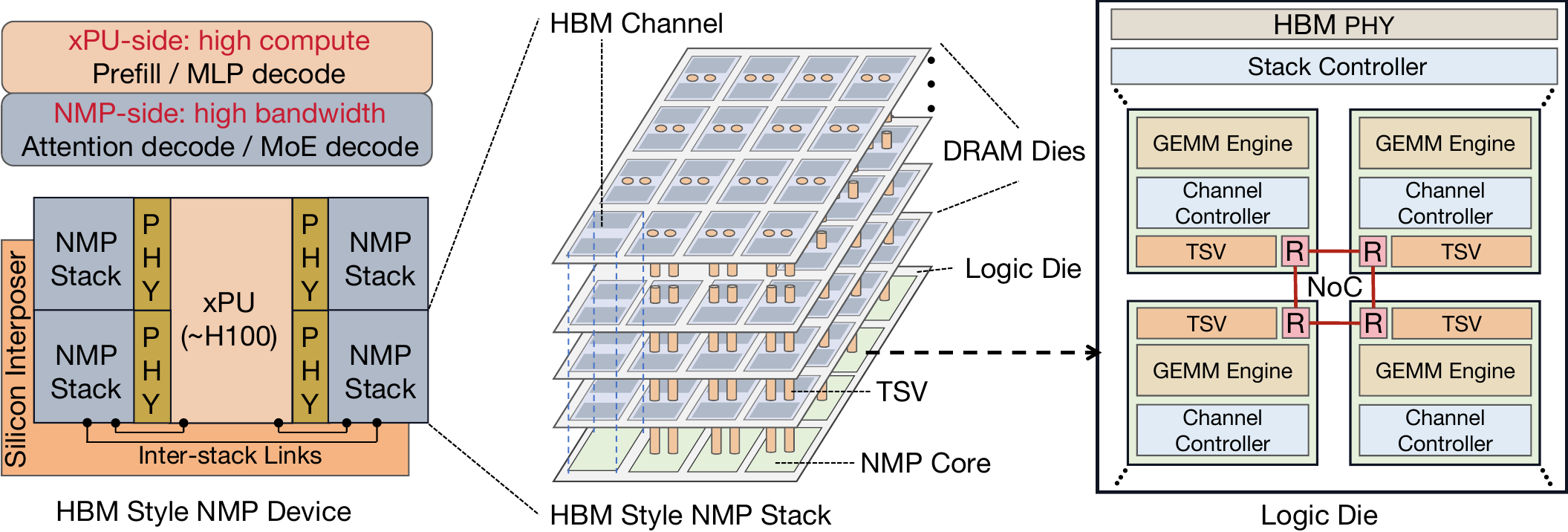}
    % \caption{Stacked DRAM NMP architecture.}
    \caption{Stacked-DRAM NMP system organization.}
    \label{fig:nmp_architecture}
\end{figure}

\begin{figure*}[tbp]
    \centering
    \includegraphics[width=0.96\textwidth]{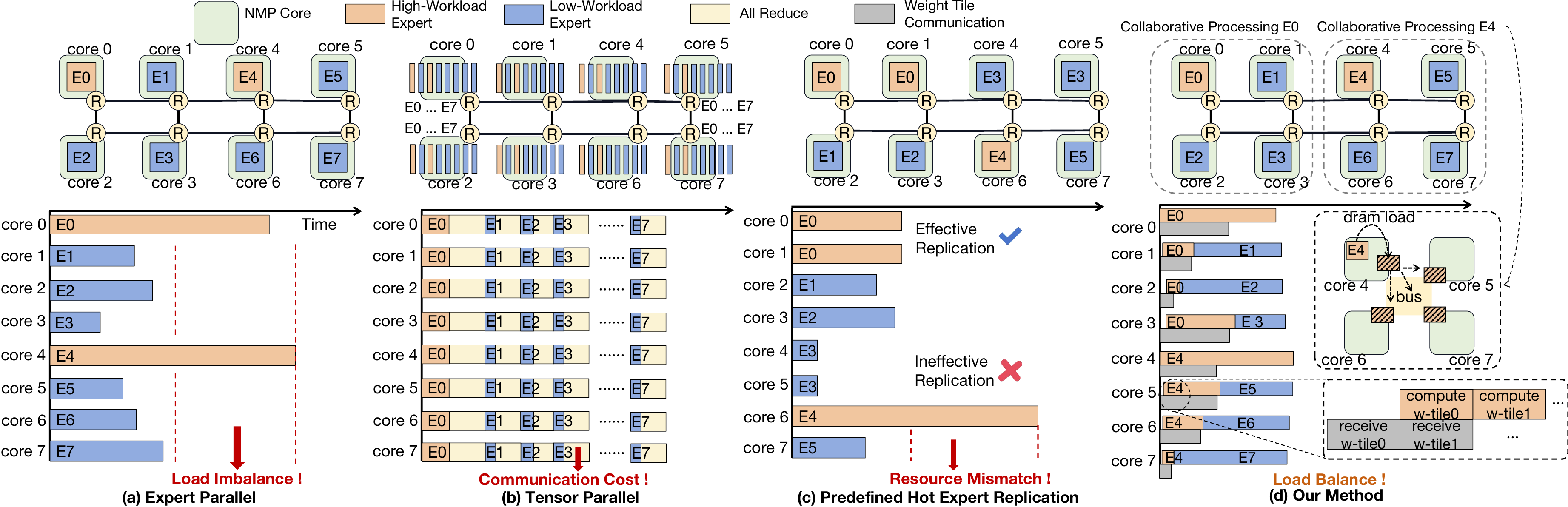}
    \caption{A comparison of prior MoE deployment strategies and our proposed method.}
    % \vspace{-0.5em}
    \label{fig:overall_compare}
\end{figure*}

% \subsection{3D-Stacked DRAM NMP Architecture: Meeting the Demands of MoE}
\subsection{3D-stacked DRAM NMP for MoE Decode}

The memory-bound nature of MoE decode is well matched to 3D-stacked Near-Memory Processing (NMP), whereas conventional GPUs/NPUs often hit the memory wall~\cite{pan2025stratum,pan2026trimoe}. In-DRAM NMP embeds lightweight logic inside DRAM banks~\cite{gu2020ipim,kwon202125,heo2024neupims,park2024attacc,gu2025pim,yang2024recpim,li2024specpim,he2025papi,lee2021hardware,li2026hardware}, but DRAM area and process technology constrain its compute capability~\cite{li2025h2}. In contrast, 3D-stacked NMP~\cite{yun2024duplex,gao2017tetris,kim2016neurocube,tian2023abndp,xie2021spacea,besta2021sisa,zhang2018graphp,zhuo2019graphq,singh2019napel,nai2017graphpim} leverages the existing base logic die beneath the stacked DRAM dies for near-memory compute. Fabricated in an advanced logic process, this die can accommodate substantially more compute engines than in-DRAM designs, while dense through-silicon vias (TSVs) provide high-bandwidth, low-latency data access. This bandwidth--compute balance matches MoE decode (Fig.~\ref{fig:moe_structure}(c)); Fig.~\ref{fig:nmp_architecture} shows the HBM-style instance evaluated here.

Recent systems such as Duplex and Stratum~\cite{yun2024duplex,pan2025stratum} adopt a heterogeneous xPU--NMP organization: the xPU executes compute-intensive operators, while NMP logic handles bandwidth-bound operators near DRAM. In our HBM-style instance, each logic-die core is tightly coupled to an HBM channel, forming a compute--memory unit; these cores communicate through a NoC. We call this unit an NMP core. Each core integrates a GEMM engine for matrix and nonlinear operations, on-die buffers, a channel controller, and a NoC router~\cite{yun2024duplex}. This tiled organization provides abundant parallelism and massive aggregate bandwidth, enabling experts to be distributed across cores and executed near data. However, realizing these benefits requires expert execution to remain near its designated memory partition.

\subsection{Parallelizing MoE on Distributed NMP Cores}\label{subsec:2.3}

Parallelizing MoE on this tiled architecture is difficult because NMP has a strong \textit{cost asymmetry}: local DRAM accesses are fast and bandwidth-rich, whereas remote accesses traverse the NoC and consume shared bandwidth. An effective design must therefore preserve locality, balance load, and control communication cost, as summarized in Fig.~\ref{fig:overall_compare}.

\begin{sloppypar}
\textbf{Expert Parallelism (EP)} maps each expert to dedicated NMP cores~\cite{lepikhin2020gshard}. EP preserves expert locality, forms larger expert GEMMs by aggregating routed tokens~\cite{liu2024deepseek3}, and keeps each token's routing fan-out independent of total expert count under fixed top-$k$~\cite{jin2025megascalemoe}. However, skewed and dynamic expert loads (Fig.~\ref{fig:expert_act_hot_diff}) turn cores hosting hot experts into stragglers.
\end{sloppypar}

\textbf{Tensor Parallelism (TP)} distributes each expert across cores and smooths its work~\cite{shoeybi2019megatron}. On 3D-stacked NMP, however, every invocation becomes a distributed execution followed by reduction-style synchronization over the mesh NoC~\cite{huang2021communication,he2025waferllm}; in our analysis, NoC overhead reaches 52\% of expert execution time at 32 cores. Weight partitioning also weakens local reuse within each core's GEMM engine. Thus, TP sacrifices the locality that benefits memory-bound decode. Restricting TP to nearby cores or hot experts in \textbf{hybrid schemes}~\cite{huang2025hd} reduces, but does not remove, the partitioning and synchronization costs. \textbf{Pipeline Parallelism (PP)} is not considered, since splitting decode batches into micro-batches weakens data reuse and introduces pipeline bubbles~\cite{huang2019gpipe}.

\textbf{Dynamic optimization on top of EP.}
Given TP's communication overhead, many systems retain EP and repair its imbalance through expert remapping, replication, or task stealing~\cite{deepseek_eplb_2025,song2024promoe,go2025moetuner,he2022fastermoe,zhai2023smartmoe,zhang2025popfetcher,yao2024exploiting,liu2025netmoe,tian2024ndpbridge,wang2021nasguard,blumofe1999scheduling,shiina2019almost}. These methods are less effective on NMP for two reasons. First, step- and layer-level variation makes hot-expert prediction difficult and limits the effectiveness of proactive reallocation; prior work reports below 50\% accuracy~\cite{fang2025klotski}. Second, weight migration and remote helper requests consume the NoC bandwidth on which NMP locality depends.

Overall, \textbf{EP is the most suitable backbone for MoE on 3D-stacked NMP}: it best matches the platform's locality-oriented cost structure. The remaining challenge is to balance load without sacrificing locality.

\subsection{Design Implications for EP Deployment}
\label{subsec:2.4}

Modern MoE models may contain 128 or more experts, while NMP exposes a fine-grained multi-core architecture, making global core-level load balance difficult. A more natural objective is to avoid localized hotspots: partition experts into groups whose aggregate loads remain balanced over time, rather than allowing multiple hot experts to cluster in the same physical region. Residual imbalance is then confined within each group, where its impact and the cost of further optimization are lower. This objective is also consistent with the general principle of multi-GPU MoE deployment~\cite{deepseek_eplb_2025}.

\section{Expert Complementarity}\label{sec:expert_complementarity}
\subsection{Domain-Specialized Routing and Expert Complementarity}\label{subsec:3.1}

As in Sec.~\ref{subsec:2.1}, MoE routing uses a learned gating network. Although MoE training includes load-balancing objectives, routed experts still specialize by domain rather than serving all inputs uniformly~\cite{dong2025domain,qiu2025demons}. This tendency strengthens in many-expert models: as shared experts absorb common patterns, routed experts become domain-specialized~\cite{yang2025qwen3,dai2024deepseekmoe}. Figure~\ref{fig:expert_act_hot_diff}(a) shows that different Qwen-MoE domains prefer different expert subsets; our profiling further finds that over 90\% of hits in math- and code-related workloads fall within the corresponding preferred subsets, consistent with Stratum~\cite{pan2025stratum}. When the request mix changes, demand shifts across these subsets, causing some expert loads to rise while others fall. We call this trace-level load cancellation \textbf{expert complementarity}. Because this structure reflects learned specialization rather than random fluctuations, grouping complementary experts can stabilize aggregate group load as targeted in Sec.~\ref{subsec:2.4}.

\subsection{Trace Construction and Metric Definition}\label{subsec:3.2}
\begin{figure}[tbp]
    \centering
    \includegraphics[width=0.9\linewidth]{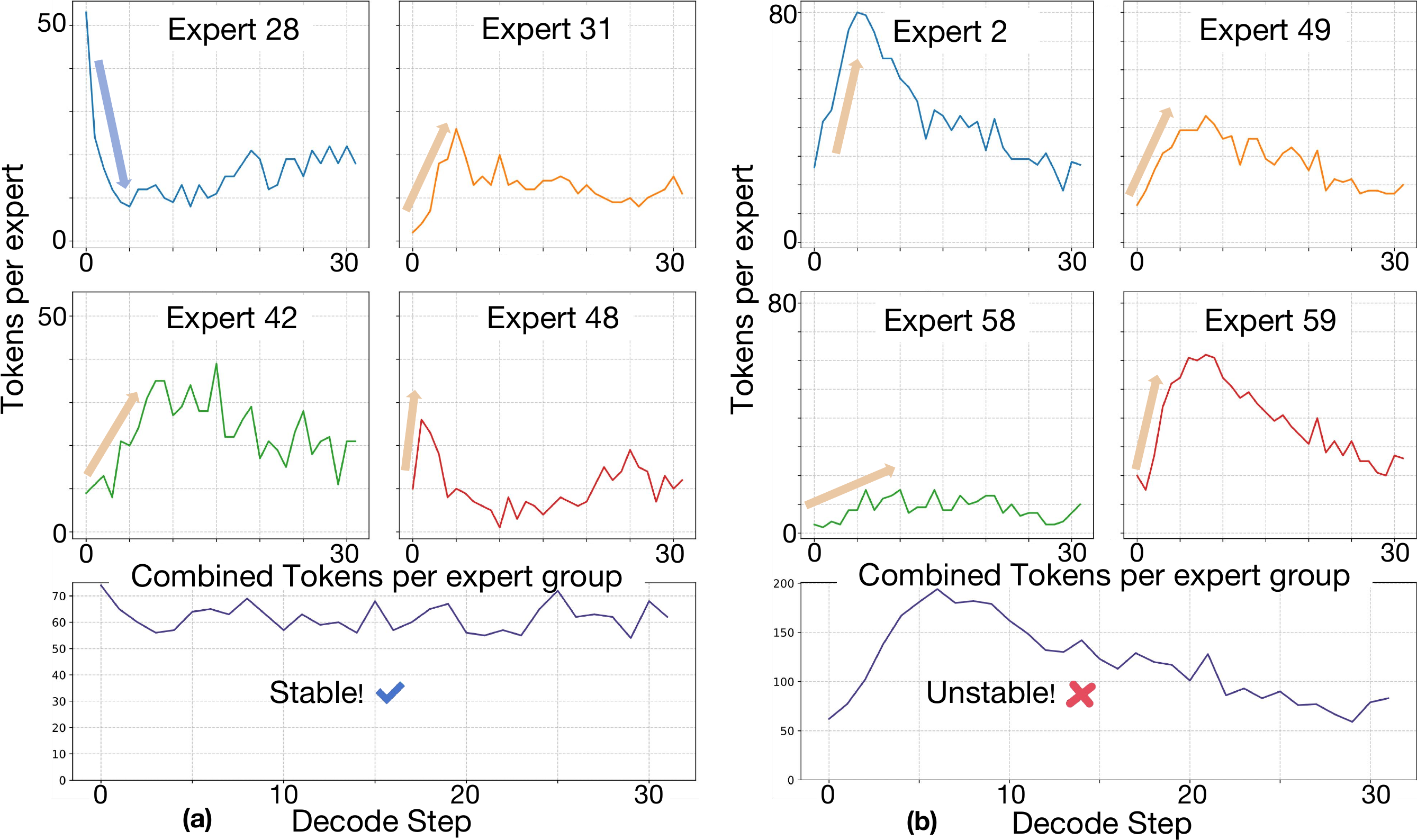}
    % \vspace{-1em}
    \caption{Complementarity examples from layer 10 of DeepSeek-V2-Lite. (a) High-complementarity experts. (b) Low-complementarity experts.}
    % \vspace{-1em}
    \label{fig:expert_comple}
\end{figure}

We operationalize complementarity with \textbf{decode-step-level expert-load traces}. For each layer, expert $e$ is represented by its routed-token count $F_{e,d}$ at step $d$ under batched serving. Figure~\ref{fig:expert_comple} shows two examples. Highly complementary traces exhibit anti-phase behavior and yield a smoother aggregate load, whereas low-complementarity traces co-vary or show no systematic cancellation.

\begin{table}[tp]
  \centering
  \scriptsize
  \setlength{\tabcolsep}{12pt}
  \renewcommand{\arraystretch}{1.0}
  \begin{threeparttable}
    \caption{Domain-preference similarity across pair types.}
    \label{tab:domain_pref_sim}
    \begin{tabular}{lccc}
      \Xhline{1.1pt}
      \textbf{Model} & \textbf{Complementary} & \textbf{Positive} & \textbf{Random} \\
      \Xhline{0.8pt}
      DeepSeek-V2-Lite  & 0.19 & 0.67 & 0.43 \\
      Qwen3-30B-A3B     & 0.25 & 0.71 & 0.46 \\
      GLM-4.5-Air       & 0.21 & 0.69 & 0.44 \\
      \Xhline{1.1pt}
    \end{tabular}
    \begin{tablenotes}[flushleft]
      \scriptsize
      \item Complementary pairs are significantly negatively correlated; positive pairs are significantly positively correlated; random pairs are sampled uniformly.
    \end{tablenotes}
  \end{threeparttable}
  % \vspace{-1em}
\end{table}

% \begin{table}[tp]
%   \centering
%   \scriptsize
%   \setlength{\tabcolsep}{3pt}
%   \renewcommand{\arraystretch}{1.0}
%   \begin{threeparttable}
%       \caption{Expert complementarity across models and datasets.}
%       \label{table:complementarity_analysis}
%       \begin{tabular}{lccc}
%         \Xhline{1.1pt}
%         & \textbf{LMSys~\cite{zheng2023lmsys}} & \textbf{ShareGPT~\cite{liyucheng_sharegpt90k_2023}} & \textbf{HumanEval~\cite{zheng2023codegeex}} \\
%         \Xhline{0.8pt}
%         DeepSeek-V2-Lite (64 experts) & 7.1 / 82.8\% / 75.7\% & 6.4 / 83.7\% / 77.1\% & 5.8 / 82.2\% / 77.3\% \\
%         Qwen3-30B-A3B (128 experts) & 10.1 / 82.5\% / 74.8\% & 10.9 / 86.0\% / 72.1\% & 12.2 / 88.3\% / 80.5\% \\
%         GLM-4.5-Air (128 experts) & 9.8 / 82.8\% / 70.6\% & 11.3 / 85.6\% / 74.5\% & 12.5 / 87.7\% / 79.4\% \\
%         \Xhline{1.1pt}
%       \end{tabular}
%       \begin{tablenotes}[flushleft]
%           \scriptsize
%           \item Entries are reported as partners / coverage / cross-dataset retention.
%       \end{tablenotes}
%   \end{threeparttable}
%   \vspace{-1em}
% \end{table}

\begin{table}[tp]
  \centering
  \begin{threeparttable}
    \caption{Expert complementarity across models and datasets.}
    \label{table:complementarity_analysis}

    \scriptsize
    \setlength{\tabcolsep}{1.5pt}
    \setlength{\arrayrulewidth}{0.3pt}
    \renewcommand{\arraystretch}{1.08}

    \begin{tabular}{@{}c|c|c|c@{}}
      \Xhline{1.1pt}

      \textbf{Model}
      &
      \textbf{LMSys~\cite{zheng2023lmsys}}
      &
      \textbf{ShareGPT~\cite{liyucheng_sharegpt90k_2023}}
      &
      \textbf{HumanEval~\cite{zheng2023codegeex}}
      \\

      \Xhline{0.8pt}

        \makecell[c]{DeepSeek (64 E)}
      &
      7.1 / 82.8\% / 75.7\%
      &
      6.4 / 83.7\% / 77.1\%
      &
      5.8 / 82.2\% / 77.3\%
      \\

      \Xhline{0.3pt}

        \makecell[c]{Qwen3 (128 E)}
      &
      10.1 / 82.5\% / 74.8\%
      &
      10.9 / 86.0\% / 72.1\%
      &
      12.2 / 88.3\% / 80.5\%
      \\

      \Xhline{0.3pt}

        \makecell[c]{GLM-4.5 (128 E)}
      &
      9.8 / 82.8\% / 70.6\%
      &
      11.3 / 85.6\% / 74.5\%
      &
      12.5 / 87.7\% / 79.4\%
      \\

      \Xhline{1.1pt}

      \multicolumn{4}{c}{%
        \scriptsize
        $E$: routed-expert count; entries are partners / coverage / cross-dataset retention.
      }
      \\

    \end{tabular}
  \end{threeparttable}

  \vspace{-1em}
\end{table}
We quantify load offsetting with Pearson correlation between expert traces: experts that consistently offset each other's token volume can smooth aggregate runtime load when grouped. Instead of using a model-independent threshold, we construct an \textit{empirical null distribution} for each model--dataset setting by repeatedly circularly shifting one expert trace relative to the other and recomputing the correlation~\cite{cohen2013statistical}. This preserves each trace's marginal load while disrupting temporal alignment~\cite{yuan2024rigorous}. A pair is \emph{significantly complementary} when its observed correlation falls below the lower $\alpha$-quantile of this null (the 5th percentile for $\alpha=0.05$). These significant pairwise relations reveal a stable complementary structure guiding expert grouping. We call such expert pairs \emph{complementary pairs}; all tests and statistics are computed independently per layer.

\subsection{Validation and Stability of Complementarity}\label{subsec:3.3}
We first test whether complementarity is consistent with domain-specialized routing. For each expert, we build a normalized preference profile from four representative datasets: HumanEval~\cite{zheng2023codegeex} for code, MATH~\cite{hendrycks2021measuring} for math, SciQ~\cite{welbl2017crowdsourcing} for science, and LogiQA~\cite{liu2020logiqa} for logic. Table~\ref{tab:domain_pref_sim} shows that complementary pairs have lower cosine similarity than positively correlated and random pairs across models, supporting our interpretation that complementarity is associated with load shifts across domain-preferred expert subsets.

We then evaluate three MoE models across three serving datasets. For each model--dataset setting, Table~\ref{table:complementarity_analysis} reports the average number of significant complementary partners per expert, the fraction of experts with at least one partner, and partner retention on the other datasets for the same model. Complementarity is prevalent: 82.2\%--88.3\% of experts have at least one partner, with 9.6 partners per expert on average. Moreover, 70.6\%--80.5\% of partners identified on one dataset remain significant in the other datasets, indicating substantial \textbf{cross-dataset stability}.

We also examine \textbf{temporal stability} using serving traces. We identify significant complementary partners from the first 256 decode steps and track their retention over the next three 256-step windows. Retention is 94.5\%, 92.7\%, and 90.8\%, respectively. Thus, although individual experts exhibit volatile step-level loads, the complementary structure remains stable over time.

\begin{figure}[tbp]
    \centering
    \includegraphics[width=1.0\linewidth]{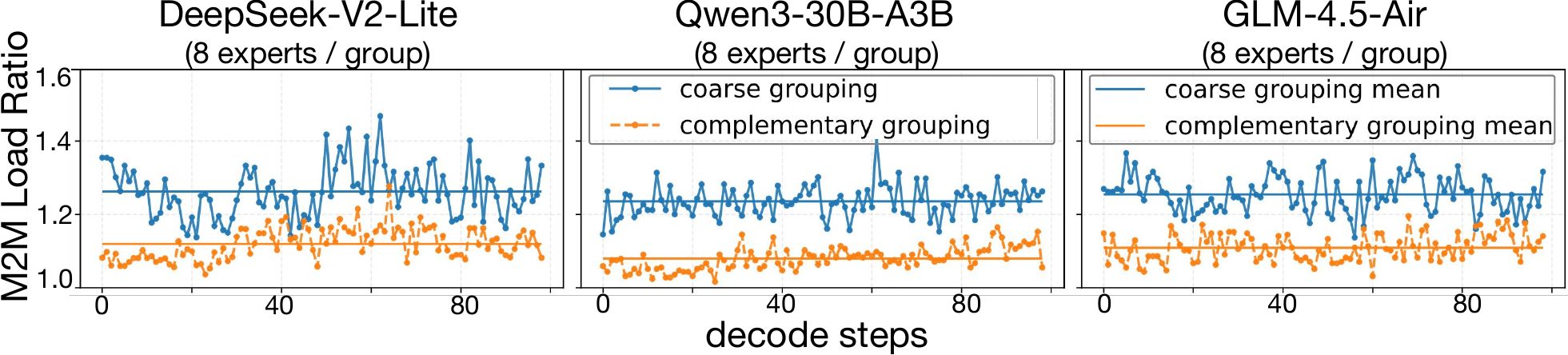}
    % \vspace{-1.5em}
    % \caption{Inter-group load balance under complementarity-guided grouping.}
    \caption{Inter-group load balance.}
    \vspace{-1em}
    \label{fig:comple_group_eva}
\end{figure}

\subsection{From Complementarity to Group-Level Load Balance}\label{subsec:3.4}
We next test whether complementarity improves load balance across expert groups. Figure~\ref{fig:comple_group_eva} compares three MoE models on LMSys under matched group sizes. We quantify per-step imbalance using the \textit{max-to-mean load ratio (M2M)}, i.e., the maximum group load divided by the mean group load; lower is better and 1 is ideal. Unlike coarse history-load-guided expert grouping~\cite{go2025moetuner,li_semantic_2026}, our method groups experts using complementarity from finer-grained decode traces, as detailed in Sec.~\ref{subsec:ILP}. Both methods use historical traces for grouping and later traces for evaluation. Across the three models, our method reduces average M2M by 17\%, keeping it near 1.1, demonstrating effective inter-group load balancing. A more comprehensive evaluation appears in Sec.~\ref{sec:algo_eva}.

The benefit is primarily group-level: complementary experts smooth aggregate load, but individual experts within a group may still exhibit transient imbalance. Under mixed-domain serving, preferred subsets can offset one another; under homogeneous serving, experts with similar preferences are less likely to cluster within the same group. Residual intra-group hotspots are handled by the lightweight hardware support in Sec.~\ref{subsec:hardware_design}. Both expert grouping and mapping rely only on decode-step load statistics, not explicit domain labels.

\begin{figure*}[t]
    \centering
    \includegraphics[width=0.95\textwidth]{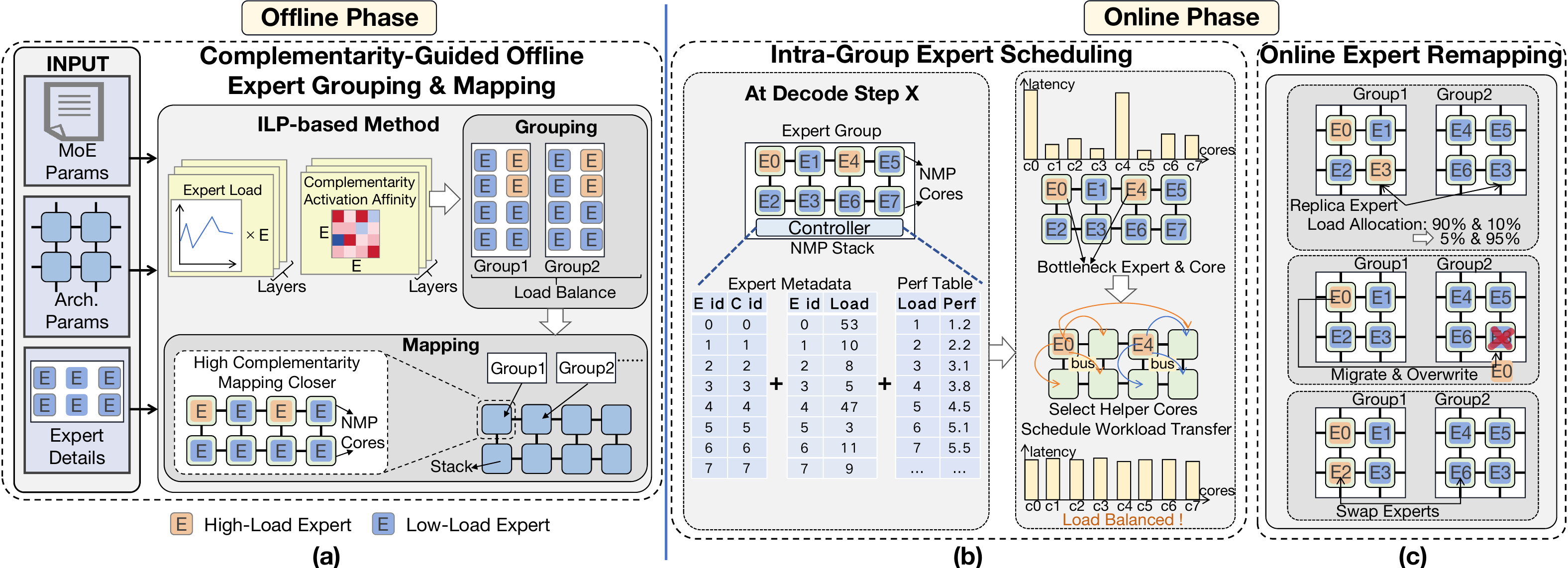}
    \caption{Software framework of 3D-MoE system: (a) Offline complementarity-guided expert grouping and mapping. (b) Runtime intra-group expert scheduling. (c) Online adaptive expert remapping.}
    \label{fig:workflow}
\end{figure*}

\section{3D-MoE System}

% TODO:关于分组的一些考量，组多大等等要说明白，类似审稿人问的那个关于HD-MoE的问题。
\subsection{System Overview}
\label{sec:system overview}
Building on the complementarity in Sec.~\ref{sec:expert_complementarity}, we propose 3D-MoE, a hardware--software co-designed system for high-throughput MoE inference on 3D-stacked NMP. 3D-MoE combines offline expert grouping and mapping, lightweight hardware support, and online scheduling and remapping.

\textit{(1) Offline Grouping \& Mapping.} Using decode-step load traces, complementarity-guided expert grouping balances aggregate loads across expert groups, while locality-aware mapping places each group's experts on nearby NMP cores to localize collaboration.

\textit{(2) Hardware Support.} Complementarity improves group-level balance but leaves transient intra-group hotspots. We therefore propose the Collaborative Memory-Network (CMN), which reads each hot expert's weight tile once from its host core's local DRAM and streams it to nearby helper cores over a reconfigured broadcast path, avoiding repeated remote reads and per-hop processing in the packet-switched NoC.

\textit{(3) Online Scheduling \& Remapping.} At runtime, intra-group scheduling selects helper cores and configures CMN for step-level hotspots, while cost-aware remapping adjusts expert replicas and placement under longer-term workload drift.

\textbf{Workflow and Scope.} 3D-MoE targets MoE layers during autoregressive decode, which account for 83--92\% of end-to-end latency. Following the cooperative xPU--NMP model~\cite{yun2024duplex}, the xPU handles prefill and compute-intensive dense operators, while NMP executes bandwidth-bound attention, gating, and MoE experts. Routing metadata then drives token dispatch, expert execution, collaboration, and output gathering on NMP. For dense models, 3D-MoE bypasses the expert-specific mechanisms and follows the baseline path~\cite{yun2024duplex,li2026hardware}.

\subsection{Complementarity-Guided Offline Expert Grouping and Mapping}
\label{subsec:ILP}

As Fig.~\ref{fig:workflow}(a) shows, 3D-MoE derives a static expert grouping and mapping plan independently for each layer. Expert grouping balances inter-group load and minimizes intra-group collaboration overhead, while physical mapping also considers token-routing locality~\cite{nie2022hetumoe}. Smaller groups simplify intra-group collaboration, but overly fine partitioning leaves less room for complementarity to smooth group load. We therefore select group size according to model and hardware constraints and keep each group within one stack whenever capacity permits. Using direct decode-step loads rather than Pearson correlation, we formulate expert grouping as an ILP and subsequent mapping as a binary quadratic placement problem.

{\setlength{\tabcolsep}{2pt}
\begin{table}[tbp]
	\caption{Grouping and mapping notation.}
	\small
    \vspace{-1em}
	\begin{center}
		\begin{tabular}{ c | p{6.3cm} }
			\hline
                \rowcolor{gray!20}
			\multicolumn{2}{c}{\textbf{Notation: for each Transformer layer}} \\
			\hline
                $\mathbb{E}$ & All experts\\
                \hline
                $\mathbb{D}$ & All decode steps\\
                \hline
                $\mathbb{G}$ & All expert groups\\
                \hline
                $\mathbb{N}$ & All NMP cores\\
                \hline
                % $R$ & Maximum number of redundant experts\\
                % \hline
                $F_{e,d}$ & Routed-token count of expert $e$ at decode step $d$\\
                \hline
                $C_{e_1,e_2,d}$ & Pairwise load-offset score of expert $e_1$ and $e_2$ at decode step $d$\\
                \hline
                $A_{e_1,e_2}$ & Co-activation affinity of experts $e_1$ and $e_2$, defined as their normalized co-activation frequency \\
                \hline
                $T_{g,d}$ / $M_{g,d}$ & Computation / Collaboration cost for expert group $g$ at decode step $d$\\
                \hline
                $B_{e,g}\in\{0,1\}$ & 1 if expert $e$ is assigned to group $g$, 0 otherwise\\ 
                \hline
                $D_{e,g}\in[0,1]$ & Workload fraction of expert $e$ assigned to group $g$ (zero if $B_{e,g}=0$)\\
                \hline
                $P_{e,n}\in\{0,1\}$ & 1 if expert $e$ is mapped to NMP core $n$\\
                \hline
                $H_{n_1,n_2}\in\mathbb{Z}_{\ge 0}$ & Communication cost between cores $n_1$ and $n_2$\\
                \hline
                
                % \rowcolor{gray!20}            \multicolumn{2}{c}{\textbf{Variables}} \\
                % \hline
                % $M_{l,i,o,n}$ & If the value is 1, it indicates that the core-$n$ is responsible for processing the $i/o$-th part of the input/output channels of the tiled layer $l$; otherwise, it is 0. \\
                % \hline
		\end{tabular}
        % \vspace{-1.5em}
		\label{table:ILP_symbol}
	\end{center}
    \vspace{-1em}
\end{table}
}

\textbf{Expert Grouping Objective.} We use a hierarchical multi-objective ILP. At decode step $d$, each group $g$ incurs computation cost $T_{g,d}$ and collaboration cost $M_{g,d}$. We first minimize the inter-group range of $T_{g,d}$ (Equation~\ref{eq:obj_comp_balance}); among layouts with comparable computation balance, we then minimize the inter-group range of $M_{g,d}$ (Equation~\ref{eq:obj_mig_balance}), followed by total collaboration cost (Equation~\ref{eq:obj_mig_total}). We show the formulation before standard linearization.

% \vspace{-1em}
\begin{equation}
\text{Minimize} \sum_{d \in \mathbb{D}}(\max_{g \in \mathbb{G}}T_{g,d}-\min_{g \in \mathbb{G}}T_{g,d})
\label{eq:obj_comp_balance}
\end{equation}

\begin{equation}
\text{Minimize} \sum_{d \in \mathbb{D}}(\max_{g \in \mathbb{G}}M_{g,d}-\min_{g \in \mathbb{G}}M_{g,d})
\label{eq:obj_mig_balance}
\end{equation}

\begin{equation}
\text{Minimize} \sum_{d \in \mathbb{D}} \sum_{g \in \mathbb{G}} M_{g,d}
\label{eq:obj_mig_total}
\end{equation}

The computation cost depends on the total tokens processed by a group. With expert replication, each replica processes a fraction of an expert's workload; thus, group computation cost sums the assigned fractions of expert token loads:

\begin{equation}
T_{g,d}=\sum_{e \in \mathbb{E}} B_{e,g} \cdot D_{e,g} \cdot F_{e,d}
\label{eq:computation_cost}
\end{equation}

Although the computation-balance objective exploits complementarity to stabilize each group's aggregate load, hot and cold experts may still coexist within a group at a given step and require local collaboration. We quantify this residual demand with a pairwise load-offset score relative to the step-average expert load, $AL_{d}=\frac{1}{|\mathbb{E}|}\sum_{e \in \mathbb{E}} F_{e,d}$. Equation~\ref{eq:complementarity} estimates the transferable workload of each pair, and Equation~\ref{eq:migration_cost} sums these scores into each group's collaboration cost. We retain the pairwise form because the subsequent mapping objective explicitly optimizes collaboration locality.

\begin{equation}
\begin{aligned}
C_{e_1,e_2,d} = 
\begin{cases}
\min\bigl(|F_{e_1,d} - AL_d|,\, |F_{e_2,d} - AL_d|\bigr) \\
\quad \text{if } (F_{e_1,d} - AL_d)(F_{e_2,d} - AL_d) < 0 \\
0 \quad \text{otherwise}
\end{cases}
\end{aligned}
\label{eq:complementarity}
\end{equation}

\begin{equation}
M_{g,d} = \sum_{e_1 \in \mathbb{E}} \sum_{{e_2 \in \mathbb{E}, e_2 > e_1}} 
B_{e_1,g} \cdot B_{e_2,g} \cdot C_{e_1,e_2,d}
\label{eq:migration_cost}
\end{equation}

\textbf{Expert Mapping Objective.} Mapping uses the pairwise scores above to place likely collaborators on nearby cores. For token routing, it also places experts frequently selected by the same token near one another; as multicast co-destinations, they can share more route segments and thus reduce redundant link traversals. Two experts are co-activated when the same token selects both under top-$k$ routing in the profiling trace; their normalized co-activation frequency defines affinity $A_{e_1,e_2}$. Under mapping $P$, $R_{e_1,e_2}=\sum_{n_1,n_2 \in \mathbb{N}}P_{e_1,n_1}P_{e_2,n_2}H_{n_1,n_2}$ is the communication cost between experts $e_1$ and $e_2$. Equation~\ref{eq:mapping_objective} combines both objectives.

% \begin{equation}
% \begin{aligned}
% \text{Minimize} \quad
% & \sum_{e_1 \in \mathbb{E}} \sum_{\substack{e_2 \in \mathbb{E} \\ e_2 > e_1}}
% \left(
% \sum_{g \in \mathbb{G}} B_{e_1,g} B_{e_2,g} \sum_{d \in \mathbb{D}} C_{e_1,e_2,d}
% + \lambda_{\text{route}} A_{e_1,e_2}
% \right)
% R_{e_1,e_2}
% \end{aligned}
% \label{eq:mapping_objective}
% \end{equation}

\begin{equation}
\begin{aligned}
\text{Minimize}\quad
& \sum_{e_1 \in \mathbb{E}}
  \sum_{\substack{e_2 \in \mathbb{E}\\ e_2 > e_1}}
  \Bigg[
    \sum_{g \in \mathbb{G}}
    B_{e_1,g} B_{e_2,g}
    \sum_{d \in \mathbb{D}}
    C_{e_1,e_2,d}
    \\[-2pt]
& \qquad\qquad
    + \lambda_{\mathrm{route}} A_{e_1,e_2}
  \Bigg]
  R_{e_1,e_2}
\end{aligned}
\label{eq:mapping_objective}
\end{equation}

Here $\lambda_{\text{route}}$ weights token-routing locality, while collaboration locality remains the primary mapping objective. The first term reduces intra-group collaboration distance; the second places co-destinations closer for multicast token dispatch. This second term affects only physical mapping, not grouping, and is a lightweight proxy rather than a full All-to-All model. Removing it increases Qwen3-30B-A3B dispatch/gather latency by 47\% on average in our evaluation.

Large experts exceeding one core's capacity are split through TP and  represented as virtual experts in the placement problem. Their all-reduce communication is included when evaluating candidate mappings, allowing the solver to trade off collaboration, token-routing locality, and TP communication.

\textbf{Constraints.} Equations~\ref{eq:replica_sum_1}--\ref{eq:core_upper} show representative constraints on workload allocation, uniform group size, and core capacity. The solver also consistent expert grouping and core placement under the stack-locality policy described above. All evaluated configurations are feasible in evaluation.

\begin{equation}
\forall e \in \mathbb{E}, \sum_{g \in \mathbb{G}}{D_{e,g}}=1
\label{eq:replica_sum_1}
\end{equation}

\begin{equation}
\forall g_1,g_2 \in \mathbb{G},\sum_
{e \in \mathbb{E}} {B_{e, g_1}}=\sum_
{e \in \mathbb{E}} {B_{e, g_2}}
\label{eq:group_same}
\end{equation}

\begin{equation}
\forall n \in \mathbb{N}, \sum_{e \in \mathbb{E}} {P_{e, n}} \le P_{\text{max}}
\label{eq:core_upper}
\end{equation}

We solve the two stages sequentially with Gurobi~\cite{gurobi}, terminating each at a 3\% relative MIP gap or after three minutes and retaining the best incumbent. In our evaluation, 68\% of optimization runs reach the 3\% target; timed-out runs have a worst final gap of 7.1\%. For each model, we profile 256 prompts over 256 decode steps from representative multi-domain traces, without domain labels. The resulting layer-wise plan is computed once and reused across workloads. Total preprocessing takes 1--3 hours per model and is amortized over deployment~\cite{zheng2022alpa}; Sec.~\ref{sec:algo_eva} shows that this fixed profiling budget captures most of the achievable grouping benefit.

\subsection{Hardware Motivation for Residual Intra-Group Hotspots}
\label{subsec:hardware}

Complementarity stabilizes aggregate group load but leaves transient imbalance among experts within a group. These localized intra-group hotspots may still require nearby underloaded cores to assist a hot expert. On 3D-stacked NMP, such collaboration is costly because the hot expert's weights remain attached to the host core.

There are three representative ways to handle such hotspots on NMP. In \emph{on-demand remote fetch}, every helper requests weight tiles from the host, repeating host service, increasing memory contention, and stalling helpers on NoC transfers. \emph{Single-read + NoC broadcast} eliminates redundant DRAM reads, but its one-to-many stream still traverses the general packet-switched network. \emph{Intra-group TP} distributes an expert across nearby cores, but every invocation requires partitioned execution and reduction, while weight partitioning weakens local reuse. Therefore, effective hotspot assistance should preserve EP residency and reduce the overhead of assistance.

We build a hotspot-expert benchmark on NMP to quantify these alternatives for 2--16 cores collaborating on one hot expert. Table~\ref{table:gap_analysis} shows that remote fetch and intra-group TP exceed the ideal latency by 29--242\% and 19--185\%, respectively; single-read NoC broadcast still adds 48\% on average. Table~\ref{tab:noc_broadcast_breakdown} explains this gap: fixed router stages contribute 63.2\% of average flit latency, versus only 7.8\% from dynamic waiting. CMN/RCL avoids repeated per-hop processing through reserved-link streaming during expert computation.

\begin{table}[tbp]
  \centering
  \caption{Characterizing hotspot collaboration on NMP.}
  \scriptsize
  \setlength{\tabcolsep}{9pt}
  \begin{tabular}{c|cccc}
    \Xhline{1.1pt}
    \#Collab Cores & 2 & 4 & 8 & 16 \\
    \Xhline{0.8pt}
    Ideal latency (ms) & 0.84 & 0.42 & 0.21 & 0.105 \\
    On-demand remote fetch (ms) & 1.08 & 0.69 & 0.45 & 0.36 \\
    Single-read + NoC broadcast (ms) & 0.94 & 0.56 & 0.33 & 0.20 \\
    Intra-group TP (ms) & 1.00 & 0.64 & 0.42 & 0.30 \\
    CMN (ms) & 0.88 & 0.46 & 0.25 & 0.14 \\
    \Xhline{1.1pt}
  \end{tabular}
  \label{table:gap_analysis}
  \vspace{-1em}
\end{table}

\begin{table}[t]
  \centering
  \caption{Flit-level breakdown of regular NoC broadcast.}
  \label{tab:noc_broadcast_breakdown}

  \scriptsize
  \setlength{\tabcolsep}{1.5pt}
  \renewcommand{\arraystretch}{1.10}

  % Vertically center the contents of tabularx columns.
  \renewcommand{\tabularxcolumn}[1]{m{#1}}

  \begin{tabularx}{\columnwidth}{
    @{}
    >{\centering\arraybackslash}m{0.10\columnwidth}
    >{\centering\arraybackslash}m{0.14\columnwidth}
    >{\centering\arraybackslash}X
    >{\centering\arraybackslash}X
    >{\centering\arraybackslash}X
    >{\centering\arraybackslash}X
    @{}
  }
    \toprule

    \makecell[c]{\textbf{Collab.}\\\textbf{cores}}
    &
    \makecell[c]{\textbf{Avg. flit}\\\textbf{lat. (cyc.)}}
    &
    \makecell[c]{\textbf{Inter-router}\\\textbf{link}}
    &
    \makecell[c]{\textbf{Inj./ej.}\\\textbf{channel}}
    &
    \makecell[c]{\textbf{Fixed router}\\\textbf{pipeline}}
    &
    \makecell[c]{\textbf{Dynamic wait}\\\textbf{/congestion}}
    \\

    \midrule

    2
    & 10.00
    & 10.0\%
    & 20.0\%
    & 60.0\%
    & 10.0\%
    \\

    4
    & 11.33
    & 11.8\%
    & 17.6\%
    & 61.8\%
    & 8.8\%
    \\

    8
    & 14.00
    & 14.3\%
    & 14.3\%
    & 64.3\%
    & 7.1\%
    \\

    16
    & 18.80
    & 17.0\%
    & 10.6\%
    & 67.0\%
    & 5.3\%
    \\

    \bottomrule
  \end{tabularx}
  \vspace{-1em}
\end{table}

These results motivate a \textbf{single-read, multi-helper} path that avoids repeated host service and packet-switched buffering, routing, and VC allocation~\cite{kim2005low,krishna2014smart}. Such a path directly matches the structured one-to-many weight stream, which conventional multihop NoCs handle inefficiently~\cite{virtualbus_noc}.

\subsection{Collaborative Memory-Network for  Intra-Group Hotspots} \label{subsec:hardware_design}

\begin{figure}[t]
    \centering
    \includegraphics[width=0.9\linewidth]{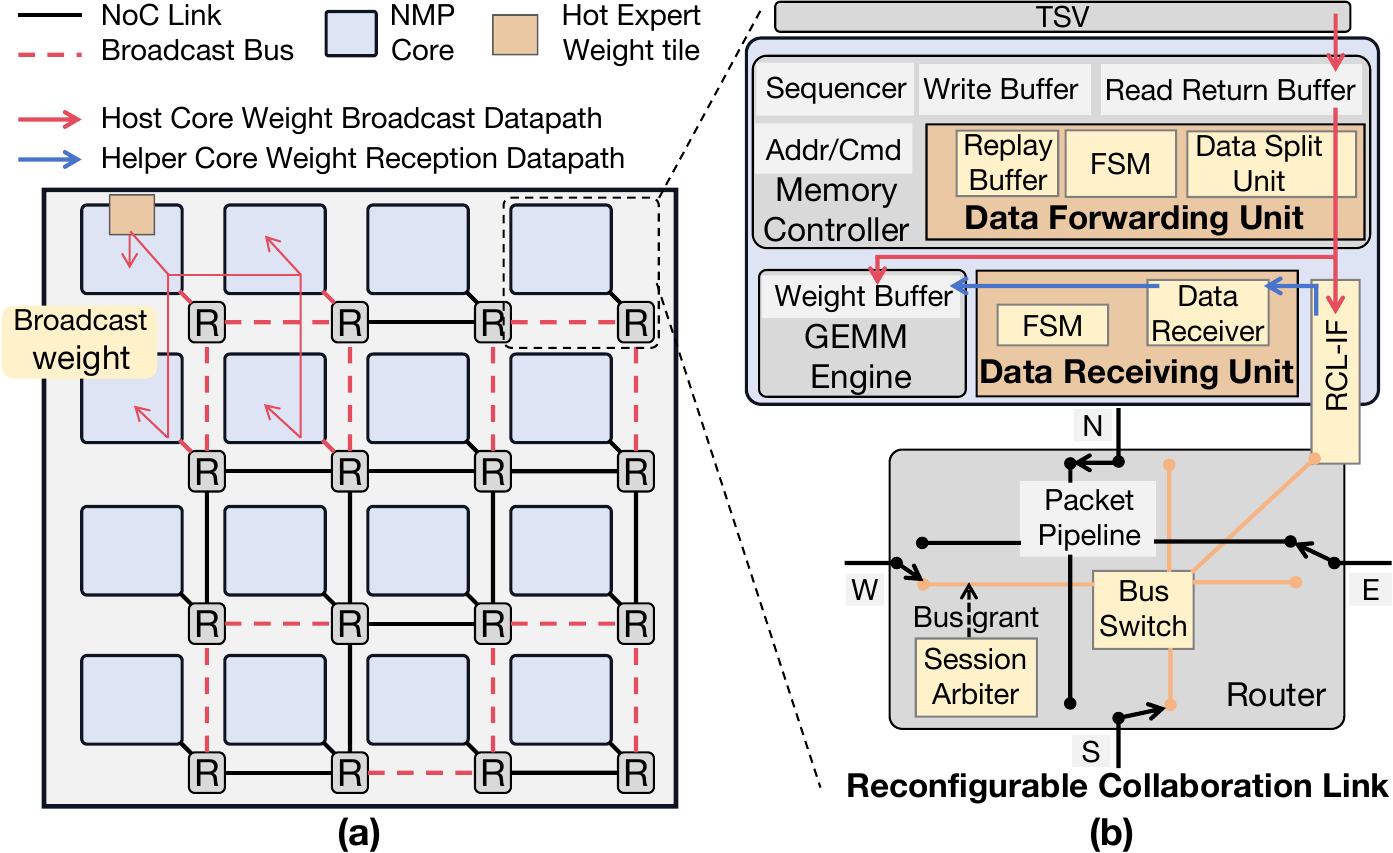}
    % \vspace{-1em}
    \caption{(a) Resource sharing via CMN. (b) CMN microarchitecture.}
    \vspace{-1em}
    \label{fig:hardware_extension}
\end{figure}

To provide the dedicated ``load once, broadcast to many'' path described above for hot-expert assistance, we propose the \textbf{Collaborative Memory-Network (CMN)}, illustrated in Fig.~\ref{fig:hardware_extension}. CMN comprises a Data Forwarding Unit (DFU), a Data Receiving Unit (DRU), and a Reconfigurable Collaboration Link (RCL) in each NMP core. During collaboration, when the host core fetches a hot expert's weight tile from its local 3D-stacked memory for computation, CMN simultaneously streams the same tile to helper cores over a temporary, bus-like local broadcast path. Thus, a single DRAM read serves all participating cores without repeated remote requests or replicated packet delivery over the regular NoC.

\textbf{Data forwarding and reception.}
The DFU and DRU implement this single-read, multi-helper path (Fig.~\ref{fig:hardware_extension}(b)). The DFU is integrated into the 3D memory controller after the Read Return Buffer. Before each collaboration task, the controller configures the DFU with the hot expert's address range and helper-core mask. When a matching DRAM read returns, the DFU forwards the tile to both the host core's local Weight Buffer and the configured RCL path. Each helper's DRU captures the incoming stream and writes it to its local Weight Buffer for immediate GEMM consumption. Helper cores therefore consume the hot expert's weights as a streamed local input rather than issuing individual remote requests.

\textbf{Reconfigurable collaboration link.}
To distribute streamed weight tiles efficiently, the RCL reuses existing NoC links in two phase-aligned modes: standard packet transport for token dispatch/gather and temporary local broadcast for hot-expert collaboration~\cite{todorov2014deterministic,manevich2009best,tahghighi2012new}. During decode, token dispatch and output gathering use packet transport, whereas residual-hotspot collaboration streams the same weight tiles from one host to several nearby helpers. Because these patterns occur in different phases, the same physical links can support both modes without concurrent contention.

Each router integrates a lightweight Bus Switch and a distributed Session Arbiter. The Bus Switch detaches selected links from the packet pipeline and reconnects them into a temporary bus-like broadcast path~\cite{zhao2012hybrid}, while the Session Arbiter grants exclusive access among collaborating hosts. Compared with replicated packets sent through the regular router pipeline, this path avoids repeated per-hop packet processing during expert computation.

Reconfiguration proceeds in three workload-aligned phases. \textbf{1) Planning:} after gating reveals the per-expert loads for the current decode step, the runtime scheduler (Sec.~\ref{sec:intra_group}) identifies residual hotspots, selects nearby helpers, and constructs collaboration domains and broadcast schedules. \textbf{2) Setup:} tokens are first delivered to the selected helpers through the regular NoC, after which the designated links enter collaboration mode. The configured path remains fixed throughout expert computation and reverts to normal NoC mode at the end of the step, amortizing the switch cost over step-level hot-expert collaboration. The RTL-derived transition takes 40 cycles: 16 cycles for session setup, an 8-cycle selected-link drain guard, and 16 cycles to restore normal NoC mode. At 650~MHz, this transition adds 61.5~ns, below 0.05\% of the CMN latency in Table~\ref{table:gap_analysis}. \textbf{3) Execution:} within each collaboration domain, the host streams each hot-expert weight tile once over the temporary broadcast path, and helpers consume it in parallel under arbiter-controlled sessions. At the end of the step, the links return to normal NoC mode.

CMN targets only residual intra-group hotspot assistance. Ordinary token dispatch, output gathering, and general inter-core communication remain on the conventional NoC. Thus, CMN does not replace the general-purpose interconnect; it is a selective fast path tailored to the structured one-host-to-many-helper pattern arising from residual intra-group hotspots. As Table~\ref{table:gap_analysis} shows, this specialized path substantially reduces the gap remaining under conventional NoC broadcast and moves collaboration latency closer to the ideal lower bound.

\subsection{Intra-group Expert Scheduling}
\label{sec:intra_group}

CMN enables low-cost collaboration, but realizing its benefits requires scheduling to identify hotspot experts per decode step and form collaboration domains and schedules. Because gating exposes per-expert loads before expert execution, the stack controller can use this window to run the latency-driven heuristic in Algorithm~\ref{alg:intra_group_scheduler}. The heuristic selects nearby helpers and constructs benefit-guided, link-disjoint collaboration domains and schedules to reduce group latency.

\begin{algorithm}[tbp]
\scriptsize
\caption{Intra-Group CMN Scheduling}
\label{alg:intra_group_scheduler}

\KwIn{Set of overloaded cores $A$; set of underloaded cores $B$;}
\KwOut{
    $D$: The final set of collaboration domains (topologies)\;
    $S$: The final ordered collaboration schedule\;
}
Sort $A$ by load (descending); Sort $B$ by load (ascending)\;
$D \gets \emptyset$; $S \gets \emptyset$\;

\BlankLine
\textcolor{magenta}{// Pair overloaded sources with underloaded helpers}\;
\ForEach{$c_{\text{src}} \in A$}{
    \ForEach{$c_{\text{dst}} \in B$}{
        $\text{Topology}_{\text{new}} \gets$ merge domains containing $c_{\text{src}}$ and $c_{\text{dst}}$\;
        
        \If{$|\text{Topology}_{\text{new}}| \le \text{MAX\_SIZE}$ \textbf{and} IsLinkConflictFree($\text{Topology}_{\text{new}}$, D)}{
            
            $\text{Plan}_{\text{new}} \gets$ GenerateOrderedScheduleFor($\text{Topology}_{\text{new}}$)\;
            \textcolor{magenta}{// Order broadcast plans within the candidate domain}\;
            
            $G \gets$ CalculateGlobalBenefit($D, S, \text{Topology}_{\text{new}}, \text{Plan}_{\text{new}}$)\;
            \textcolor{magenta}{// Estimate makespan reduction after CMN cost}\;
            
            \If{$G > 0$}{
                $D \gets$ UpdateDomains($D, \text{Topology}_{\text{new}}$)\;
                $S \gets$ UpdateSchedules($S, \text{Plan}_{\text{new}}$)\;
                Update loads of all involved cores\;
            }
        }
    }
}
\vspace{-1em}
\end{algorithm}

As Fig.~\ref{fig:workflow}(b) shows, expert metadata converts per-expert token counts into estimated core latencies through a pre-measured performance table~\cite{ke2022hercules,xie2021apollo}. Based on this latency profile, the scheduler starts from the most overloaded source core and incrementally expands its broadcast domain with underloaded helpers, forming candidate collaboration domains $\text{Topology}_{\text{new}}$. This allows part of the hot expert's routed tokens to be reassigned to the newly added helper, while CMN broadcasts the expert's weights from the source. Once the current hotspot is relieved, the scheduler proceeds to the next most overloaded core. Each candidate is limited to at most 16 cores within one stack and must be link-disjoint from active domains. The scheduler greedily retains beneficial candidates after accounting for CMN collaboration cost and generates the corresponding ordered collaboration schedule.

This procedure runs on the stack controller with compact metadata. Each expert group spans only 8--16 cores in our settings, keeping metadata small. A compact performance table covering up to 2048 tokens per expert supports the evaluated batched-decode settings with about 4\,KB of SRAM.

\subsection{Online Expert Remapping}
\label{subsec:online remapping}

Although offline grouping and mapping provide a strong initial layout, MoE workload drift can create imbalances and reduce performance by up to 15\% in our evaluation. 3D-MoE therefore uses cost-aware online expert remapping that combines frequent low-cost adjustments with selective higher-cost reconfigurations.

Online remapping uses metadata maintained by the stack controller to predict expert loads. The metadata records expert-to-core mappings, replica locations, and per-step workloads. For each expert prediction, we maintain an exponential moving average (EMA), $\text{EMA}_e(t)=\alpha F_e(t)+(1-\alpha)\text{EMA}_e(t-1)$, where $F_e(t)$ is expert $e$'s load at step $t$. The smoothing parameter $\alpha$ weights the latest observation against historical trends; we set $\alpha=0.3$ based on profiling-trace prediction accuracy. EMA remains responsive to recent changes while suppressing noise, helping identify sustained load drift.

\textbf{Multi-Tiered Adjustment.} Based on migration cost, we use three adjustment tiers (Fig.~\ref{fig:workflow}(c)). Tier-1 redistributes load fractions among existing expert replicas without moving weights. Tier-2 adds a replica for an overloaded expert and removes a redundant replica from an underutilized one; the new replica overwrites the freed space and requires one weight transfer. Tier-3 swaps experts across groups to correct severe inter-group imbalance, requiring two weight migrations.

\textbf{Cost-Aware Trigger.} Migration-free Tier-1 runs before every MoE layer. Tier-2 and Tier-3 candidates are checked every eight decode steps or earlier when sustained imbalance is detected. The eight-step interval is selected from profiling traces. Sustained imbalance is detected when EMA-predicted group-level M2M exceeds the profiled steady-state level by 20\% for three consecutive decode steps. At each check, the controller estimates candidate benefits from EMA-predicted loads, subtracts migration costs based on moved expert-weight volume, transfer distance, and available bandwidth, and executes the highest-benefit action only if net benefit is positive. Weight transfers are executed asynchronously and overlap with computation. On Qwen3-30B-A3B, each layer averages 24 weight-moving remappings over 256 decode steps; 91\% of transfer time is hidden by xPU dense execution or NMP idle windows, leaving 0.87\% exposed end-to-end overhead.

\section{Evaluation}
\subsection{Experimental Setup}

\textit{1) Target NMP System:} We evaluate all NMP schemes on the same HBM-style heterogeneous xPU--NMP platform. Each NMP device is a single package comprising one H100-class xPU die and four 16-GB NMP stacks interconnected through a silicon interposer~\cite{yun2024duplex,pan2025stratum}. HBM PHYs provide the xPU--stack interfaces, while the four stack logic dies form a $2\times2$ interposer mesh~\cite{shao2019simba}. Based on model size, we select NMP-S/M/L with 1/2/4 NMP devices connected by an NVLink-like fabric. Each NMP stack follows Duplex's NMP compute capability and internal DRAM bandwidth for MoE decoding~\cite{yun2024duplex}; NMP-local bandwidth refers to the internal DRAM-to-logic path rather than the xPU-visible HBM interface. Each stack's logic die uses a $4\times4$ mesh NoC with 512-bit links~\cite{jiang2025ndpage}. Table~\ref{table:hardware_spec} lists the parameters.

\begin{table}[tbp]
\centering
\caption{NMP system configurations.}
\vspace{-0.5em}
\label{table:hardware_spec}
\scriptsize
\setlength{\tabcolsep}{2pt}
\renewcommand{\arraystretch}{1.0}

\begin{tabularx}{\columnwidth}{
@{}
>{\raggedright\arraybackslash}p{1.6cm}
X
@{}
}
\toprule
\textbf{Level} & \textbf{Configuration} \\
\midrule

NMP-S/M/L
& 1/2/4 NMP devices; 64/128/256 GB total capacity \\
\specialrule{0.3pt}{0pt}{0pt}

NMP device
& \begin{minipage}[t]{\linewidth}
  1 H100-class xPU die + 4 NMP stacks on one interposer\\[0pt]
  xPU: 0.99 PFLOPS; aggregate xPU--stack HBM BW: 2.68 TB/s
  \end{minipage} \\
\specialrule{0.3pt}{1.5pt}{0pt}

NMP stack
& 16-GB 8-hi DRAM; 16 channels;
16 NMP cores, $4\times4$ mesh \\
\specialrule{0.3pt}{0pt}{0pt}

NMP core
& 650 MHz; 1.33 TFLOPS;
166.4 GB/s local DRAM BW; 64-KB buffer \\
\specialrule{0.3pt}{0pt}{0pt}

Interconnect
& \begin{minipage}[t]{\linewidth}
  Intra-stack: 512-bit, 2 cycles/hop; Inter-stack: 128 GB/s/direction\\[0pt]
  Inter-device: NVLink-like, 450 GB/s/direction 
  \end{minipage} \\

\bottomrule
\end{tabularx}
\vspace{-1.1em}
\end{table}

\begin{table}[t]
\centering
\caption{MoE model configurations.}
\vspace{-0.6em}
  \scriptsize
  \setlength{\linewidth}{0pt}   % 列间距，可按需微调
  \renewcommand{\arraystretch}{1}
  \begin{tabularx}{\columnwidth}{@{}%
      >{\centering\arraybackslash}p{2.4cm}!{\vrule\vrule}%
      c!{\vrule\vrule}%
      c|c|c|c%
      @{}}
    \Xhline{1.1pt}
    \textbf{Model} & \textbf{Param} & $\mathbf{N_{ex}}$ & \textbf{top-$k$} & \textbf{GPU Cfg} & \textbf{NMP Cfg} \\
    \Xhline{0.8pt}
    DeepSeek-V2-Lite~\cite{liu2024deepseek2}             & 16B & 64  & 6 & 1$\times$H100 & NMP-S \\
    Qwen3-30B-A3B~\cite{yang2025qwen3}                   & 31B & 128 & 8 & 2$\times$H100 & NMP-M \\
    GLM-4.5-Air~\cite{team_glm-45_2025}                  & 106B & 128 & 8 & 4$\times$H100 & NMP-L \\
    \Xhline{1.1pt}
  \end{tabularx}
  \begin{tablenotes}[flushleft]
      \scriptsize
      \item DeepSeek-V2-Lite/GLM-4.5-Air include 2/1 additional shared experts, respectively.
    \end{tablenotes}
  \vspace{-1.5em}
   \label{table:model_spec}
\end{table}

\textit{2) System Models:} Our evaluation uses a cycle-accurate simulator built on Ramulator 2.0~\cite{luo2023ramulator} and BookSim~\cite{jiang2013detailed}. We extend Ramulator to model HBM3 timing~\cite{JEDEC_HBM3_2022} and NMP-core memory accesses, and extend BookSim to model packet-switched intra- and inter-stack traffic and stack-local RCL collaboration. Specifically, the RCL extension reserves selected intra-stack links at the phase level while retaining BookSim's topology and link parameters, and adds RTL-derived setup and drain costs. Computation is modeled at tile granularity: GEMM and reduction tiles are scheduled when inputs are ready, with compute, memory access, and communication overlapped when dependencies permit. We implement CMN in RTL and synthesize it with Synopsys Design Compiler in 7~nm technology~\cite{clark2016asap7} to estimate area and peak-power overhead. For end-to-end results, dense operators executed on the xPU use GPU-profiled latency, which is combined with the NMP simulator for attention and MoE execution.

\textit{3) Baselines:} We compare 3D-MoE with GPU reference systems and representative NMP deployment strategies. Every configuration in Table~\ref{table:model_spec} accommodates the model and its KV cache. GPU baselines are evaluated with vLLM 0.8.1~\cite{kwon2023efficient}. All NMP baselines use the same hardware configuration as 3D-MoE: static Expert Parallelism (\textbf{EP}) and Tensor Parallelism (\textbf{TP}); \textbf{HD-MoE}, a representative NMP-oriented hybrid strategy that applies TP to hot experts and EP to cold experts through offline analysis~\cite{huang2025hd}; EPLB-based predictive replication and mapping (\textbf{PS})~\cite{deepseek_eplb_2025}; and dynamic work stealing (\textbf{WS}).

\begin{sloppypar}
\textit{4) Workloads and Evaluation Metrics:} We construct serving workloads using real-world prompts from LMSys~\cite{zheng2023lmsys}, ShareGPT~\cite{liyucheng_sharegpt90k_2023}, MATH~\cite{hendrycks2021measuring}, and HumanEval~\cite{zheng2023codegeex}. We evaluate batch sizes 32--256 with fixed lengths (512 or 2048 tokens) and irregular sequence lengths. Each model's offline grouping/mapping plan is solved once from the representative profiling trace described in Sec.~\ref{subsec:ILP} and is reused in evaluations. We report MoE decode-layer latency, M2M ratio, compute resource utilization, and end-to-end throughput. All NMP schemes share the cooperative xPU--NMP attention path; 3D-MoE changes only the NMP execution path for token dispatch, expert execution, collaboration, and output gathering.
\end{sloppypar}

\subsection{Generalization of Offline Expert Grouping}\label{sec:algo_eva}

\begin{figure}[tbp]
    \centering
    \includegraphics[width=0.92\linewidth]{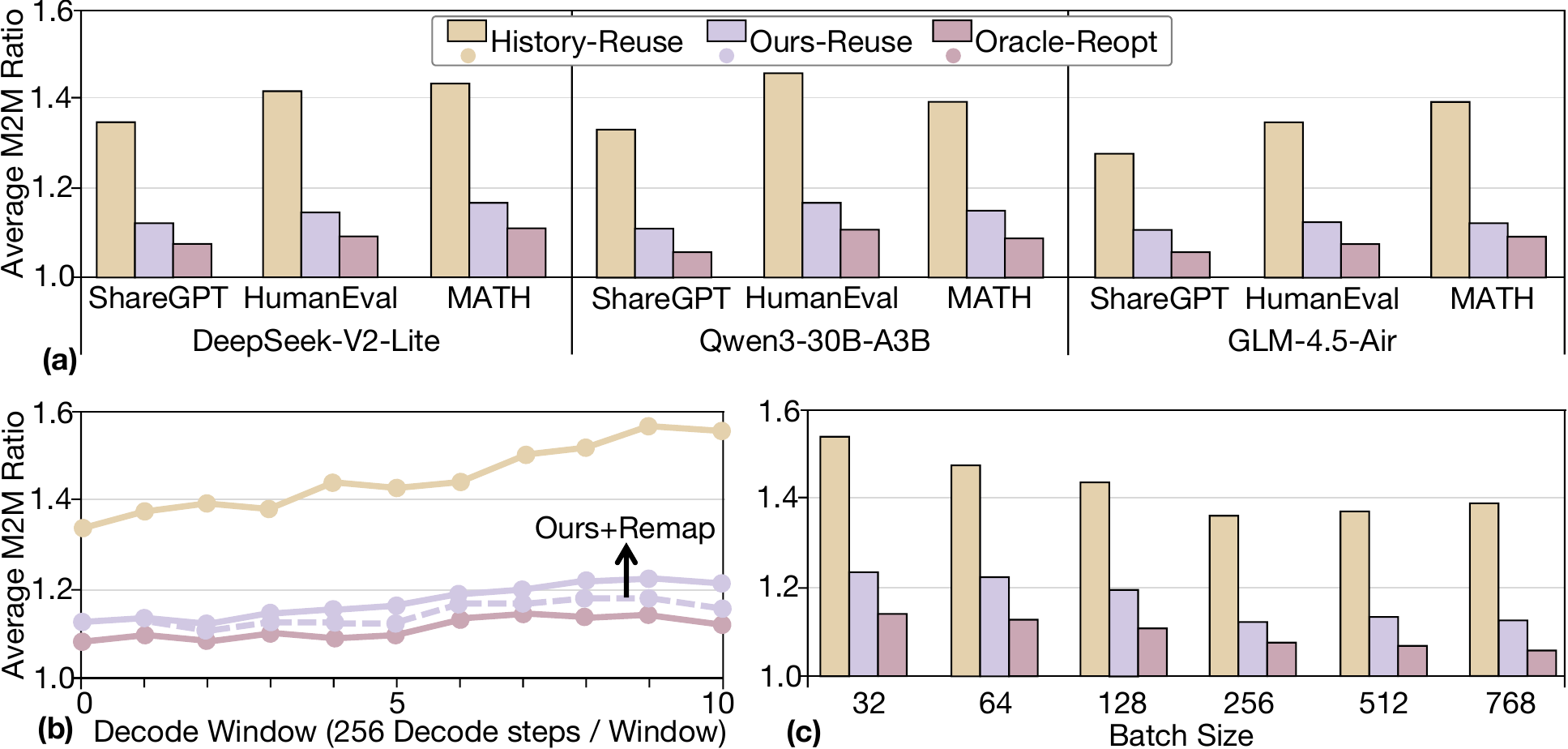}
    \vspace{-0.5em}
    \caption{Generalization of one-time offline expert grouping. (a) Cross-dataset generalization. (b) Cross-window stability; the dashed curve denotes Ours+Remap. (c) Cross-batch stability. (b) and (c) use Qwen3 on ShareGPT.}
    \vspace{-1em}
    \label{fig:algo_eva}
\end{figure}

\begin{figure*}[t]
    \centering
    \includegraphics[width=0.9\textwidth]{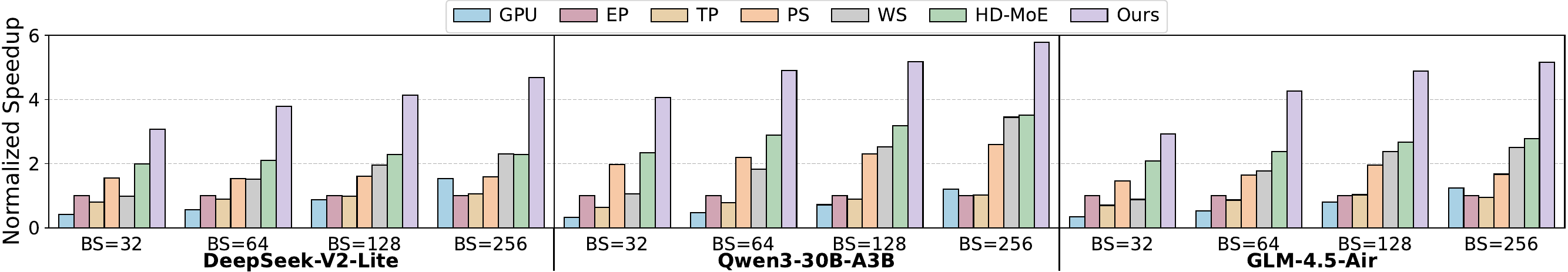}
    \vspace{-0.5em}
    \caption{MoE layer decode speedup of all schemes relative to the EP baseline under varying batch sizes.}
    % \vspace{-1em}
    \label{fig:moe_speedup}
    % \vspace{-1em}
\end{figure*}

\begin{figure*}[t]
    \centering
    \includegraphics[width=0.9\textwidth]{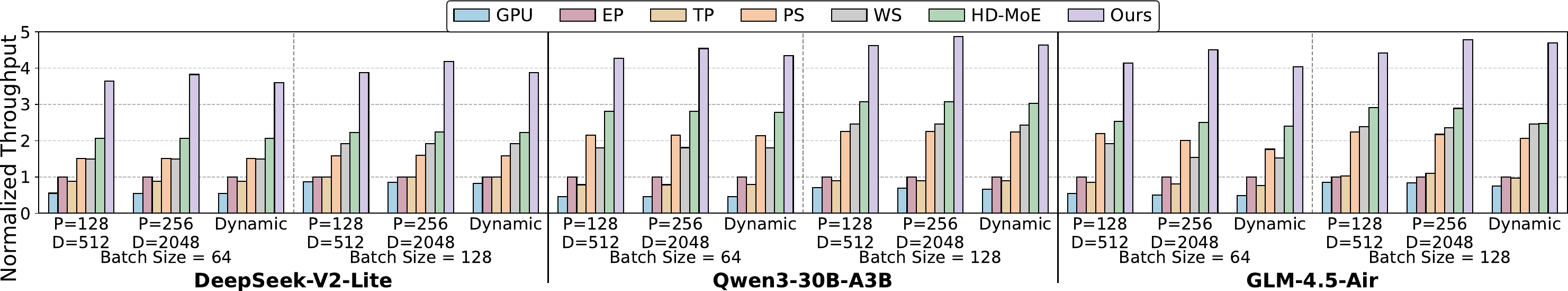}
    \vspace{-0.5em}
    \caption{End-to-end throughput of all schemes relative to the EP baseline.}
    % \vspace{-1em}
    \label{fig:moe_throughput}
    \vspace{-1em}
\end{figure*}

% As 3D-MoE's foundation, we evaluate our offline expert grouping: specifically, whether it improves inter-group balance over history-guided methods and remains robust across dynamic serving conditions. For each model, we extract 256 prompts from LMSys and profile 256 decode steps to solve the grouping plan once. We compare three settings: \textbf{History-Reuse}, a MoETuner-based grouping baseline built from coarser-grained historical load statistics~\cite{go2025moetuner}; \textbf{Ours-Reuse}, our decode-step complementarity-guided grouping solved once and then reused; and \textbf{Oracle-Reopt}, which re-solves the grouping per condition and serves as an upper bound.

We first test whether one-time offline grouping improves inter-group balance and remains robust to serving dynamics. For each model, we solve one plan from 256 LMSys prompts over 256 decode steps. The ILP uses only per-step expert loads, without domain labels, to identify complementary patterns. We compare \textbf{History-Reuse}, a MoETuner-based plan using coarse historical loads~\cite{go2025moetuner}; \textbf{Ours-Reuse}, our complementarity-guided plan; and \textbf{Oracle-Reopt}, an upper bound that re-solves the plan for each condition.

% Figure~\ref{fig:algo_eva}(a) evaluates \textbf{cross-dataset reuse}. We test the fixed grouping plan on three held-out workloads: ShareGPT, HumanEval, and MATH. Ours-Reuse achieves lower group-level M2M than History-Reuse, reducing average M2M by 20\%--24\% and closing 86\% of the gap to Oracle-Reopt. This shows that decode-step complementarity not only yields better grouping quality than history-guided methods, but also generalizes well to unseen workloads.

Figure~\ref{fig:algo_eva}(a) evaluates \textbf{cross-dataset reuse}. We test the fixed grouping plan on three held-out workloads: ShareGPT, HumanEval, and MATH. Ours-Reuse lowers group-level M2M by 20\%--24\% over History-Reuse and closes 86\% of the gap to Oracle-Reopt, showing that trace-level complementarity not only yields better grouping quality than history-guided methods, but also generalizes beyond the profiling workload without domain labels.

Figure~\ref{fig:algo_eva}(b) evaluates cross-window reuse by reporting the average M2M of the fixed grouping over successive 256-step windows from the same serving stream. Ours-Reuse degrades by only 9\% across windows and consistently outperforms History-Reuse. Ours+Remap further reduces later-window M2M, showing that online remapping effectively corrects residual long-term drift. Under an abrupt Qwen3-30B-A3B code-to-math transition, our plan lowers peak M2M to 1.21 versus History-Reuse's 1.62, and remapping reaches 1.15 within 16 steps. Figure~\ref{fig:algo_eva}(c) evaluates cross-batch reuse: Ours-Reuse remains stable from batch size 32 to 768, reducing M2M by 22\% over History-Reuse while staying within 25\% of Oracle-Reopt. Thus, a single offline plan generalizes across datasets, time windows, and batch sizes, while remapping corrects longer-term drift.

\subsection{3D-MoE Performance}
\begin{figure}[tbp]
    \centering
    \includegraphics[width=0.86\linewidth]{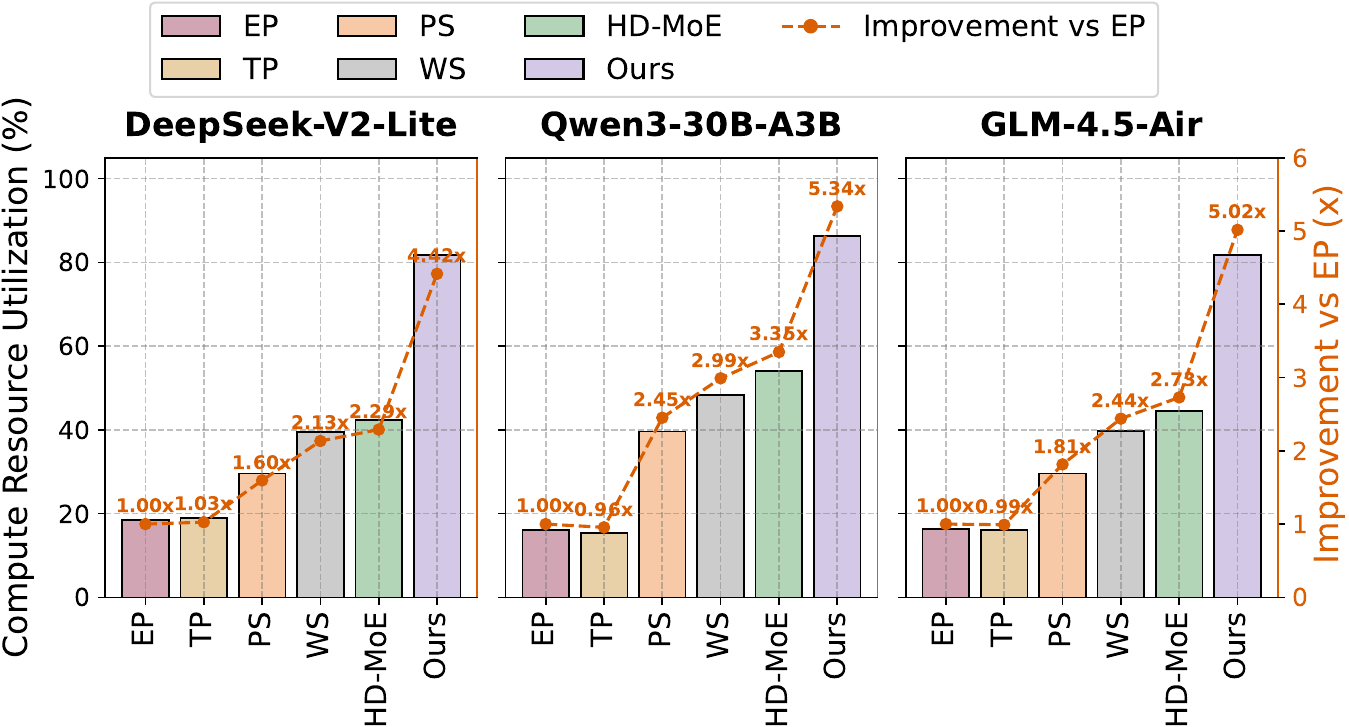}
    \vspace{-0.7em}
    \caption{Average compute resource utilization.}
    \vspace{-0.7em}
    \label{fig:avg_utilization}
\end{figure}

\textit{1) MoE Layer Decode Performance:} 3D-MoE targets the primary bottleneck in MoE inference: the expert decode phase. Figure~\ref{fig:moe_speedup} shows that 3D-MoE consistently reduces latency, with average speedups of 5.87$\times$, 4.63$\times$, and 1.72$\times$ over GPU, EP, and HD-MoE, respectively. This gain stems from 3D-MoE's highly efficient use of NMP compute resources. 3D-MoE achieves 83\% average utilization, more than 4.25$\times$ that of static deployment strategies (Fig.~\ref{fig:avg_utilization}). This highlights key shortcomings of existing schemes: EP/TP cannot adapt to dynamic workloads or incur communication overhead; PS misallocates resources due to inaccurate predictions; WS relies on costly general-purpose NoC collaboration; and HD-MoE incurs communication overhead from applying global TP to hot experts. By combining balanced placement with low-cost local collaboration, 3D-MoE achieves 1.93$\times$, 1.59$\times$, and 1.84$\times$ the utilization of the best NMP baseline on DeepSeek, Qwen, and GLM, respectively.

\textit{2) End-to-End Performance:} Across all workloads in Fig.~\ref{fig:moe_throughput}, 3D-MoE outperforms the strongest NMP baseline by 1.58$\times$--1.86$\times$, confirming that MoE-layer gains translate to system throughput. For instance, on the Qwen model, the 5.34$\times$ improvement in MoE-layer resource utilization over EP baseline efficiently yields 4.71$\times$ higher end-to-end throughput.

Compared with the GPU, 3D-MoE achieves speedups of 5.54$\times$, 7.88$\times$, and 6.95$\times$ across the three models. The GPU is severely hindered by irregular memory accesses from sparse expert activations, which makes it memory-bound. In contrast, 3D-MoE unlocks the NMP architecture's potential and converts NMP's bandwidth advantage into end-to-end gains by mitigating the load imbalance and communication overhead.

\begin{figure}[tbp]
    \centering
    \includegraphics[width=0.98\linewidth]{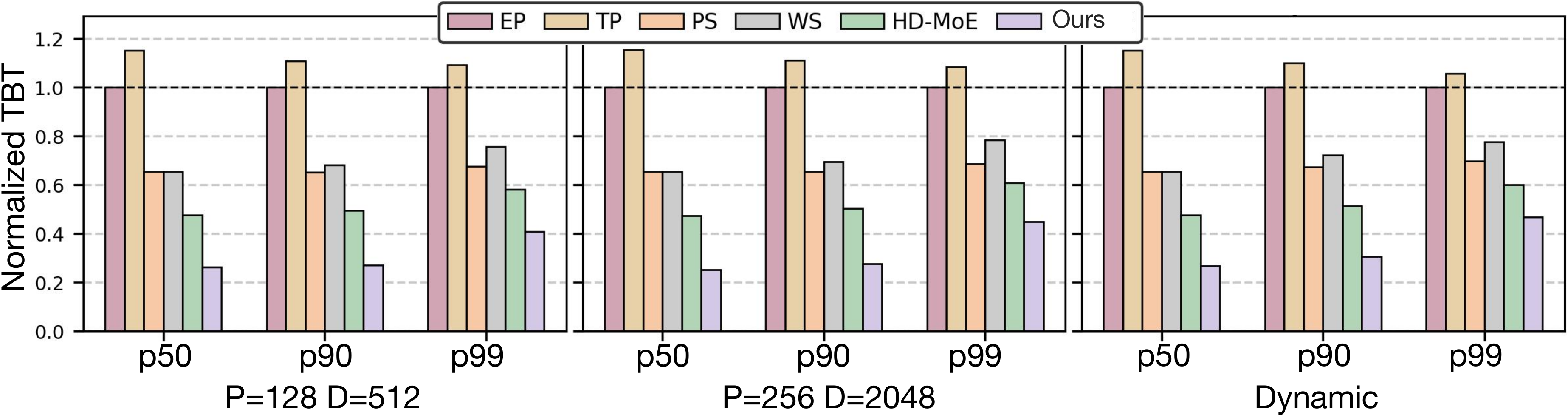}
    \vspace{-0.7em}
    \caption{TBT latency distribution on DeepSeek-V2-Lite.}
    % \vspace{-0.5em}
    \label{fig:tbt}
\end{figure}

\textit{3) Decode-Token Latency:} Because 3D-MoE targets decode, Fig.~\ref{fig:tbt} reports normalized TBT rather than TTFT for DeepSeek-V2-Lite at batch size 64. It reduces p50/p90/p99 TBT over HD-MoE by 45.3\%/43.7\%/26.0\%, showing that complementarity-guided grouping and local collaboration improve tail decode-token latency from hot-expert imbalance.

\textbf{Impact of Output Length and Batch Size.} First, 3D-MoE's speedup over the GPU baseline grows with output sequence length, as our optimizations target the autoregressive decode phase, which increasingly dominates latency in long-generation tasks. Second, 3D-MoE's advantage over other NMP baselines widens as batch size increases. This is because larger batches exacerbate expert load imbalance, an issue our complementarity-based scheduling effectively mitigates for large-scale serving scenarios.

\subsection{Ablation Study}\label{sec:ablation}

To isolate each component's contribution, we progressively enable 3D-MoE from \textbf{Random}, which uses random mapping and NoC-based collaboration: \textbf{+G/M-Coarse} groups and maps with coarse-grained historical loads; \textbf{+G-Steps} groups with decode-step loads but maps randomly; \textbf{+G/M-Steps} enables our full fine-grained grouping and mapping; \textbf{+Hardware} adds dedicated low-overhead collaboration; and \textbf{+Remapping} completes the system with online expert remapping.

\begin{figure}[tbp]
    \centering
    \includegraphics[width=0.9\linewidth]{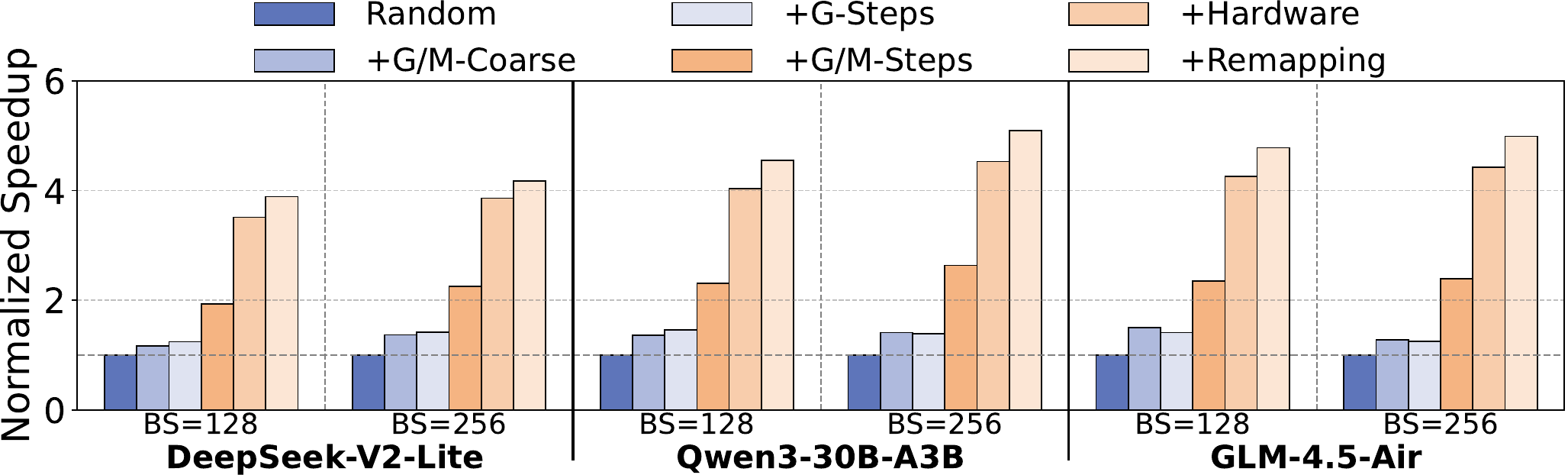}
    \vspace{-0.7em}
    \caption{Ablation of 3D-MoE components.}
    \vspace{-0.7em}
    \label{fig:ablation}
\end{figure}

Figure~\ref{fig:ablation} shows clear stepwise gains. \textbf{+G/M-Coarse} achieves a 1.35$\times$ speedup over Random by mitigating static placement mismatches. \textbf{+G/M-Steps} achieves 1.72$\times$ and 1.70$\times$ speedups over \textbf{+G/M-Coarse} and \textbf{+G-Steps}, demonstrating that effective complementarity-guided mapping requires fine-grained decode-step statistics rather than coarse historical averages. \textbf{+Hardware} provides a 1.77$\times$ speedup, validating low-overhead local collaboration. Finally, \textbf{+Remapping} adds another 1.11$\times$ under dynamic workloads by adapting the offline grouping and mapping plan to longer-term drift.

\subsection{Sensitivity Study}

\begin{figure}[tbp]
    \centering
    \includegraphics[width=0.92\linewidth]{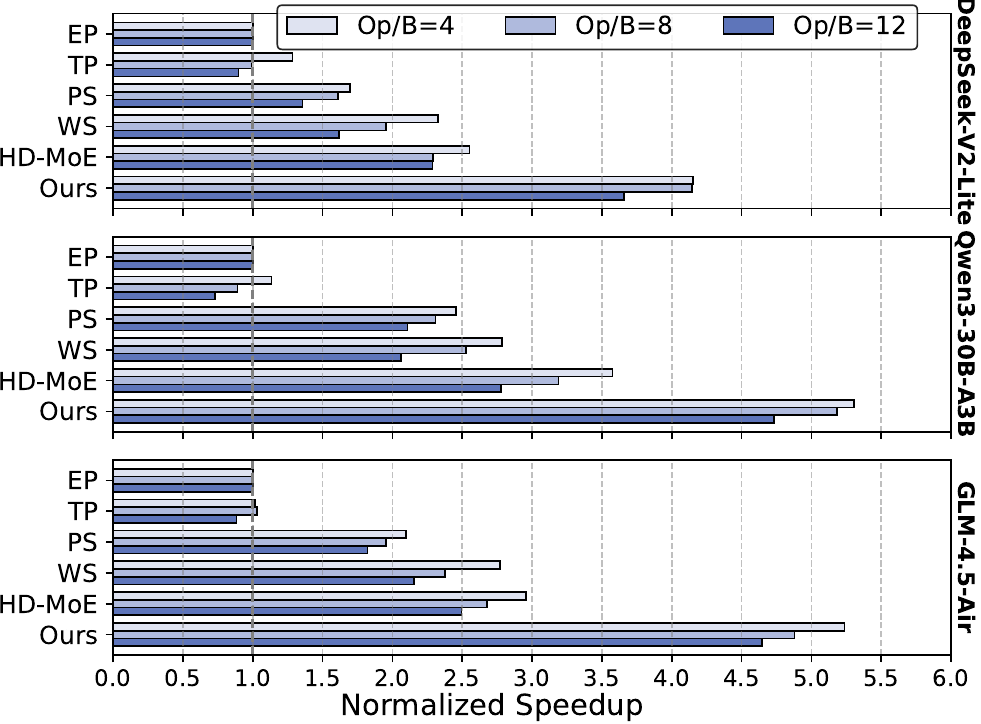}
    \vspace{-0.7em}
   \caption{Normalized MoE layer decode speedup at different Op/B ratios.}
    % \vspace{-0.2em}
    \label{fig:sensitivity}
\end{figure}

The NMP compute-to-bandwidth ratio is a key hardware design parameter. We vary compute provisioning at fixed memory bandwidth, producing machine-balance points from 4 to 12 peak Op/B. Figure~\ref{fig:sensitivity} shows that 3D-MoE consistently outperforms the best baseline by 1.59$\times$--1.86$\times$ across these design points. Compared with EP, its speedup decreases modestly from 4.9$\times$ at 4 peak Op/B to 4.34$\times$ at 12. This indicates that when compute is relatively underprovisioned, hotspot experts dominate performance and benefit more from our low-overhead collaboration support. As compute capability increases, part of the imbalance penalty is naturally hidden, but 3D-MoE remains consistently beneficial.

\begin{figure}[t]
    \centering
    \includegraphics[width=0.95\linewidth]{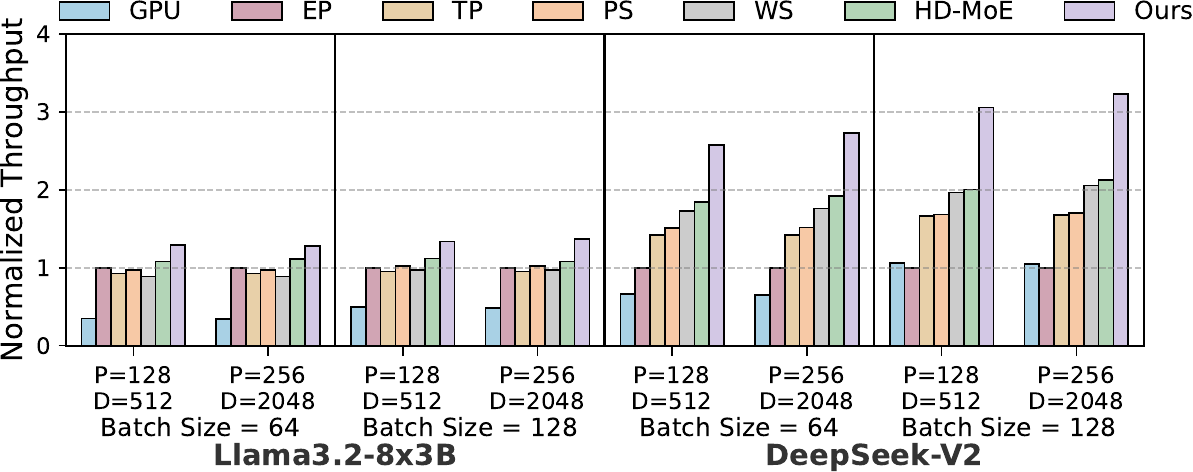}
    \vspace{-0.7em}
    \caption{Normalized throughput for scale-out and low-expert-count workloads.}
    \vspace{-0.7em}
    \label{fig:scalability}
\end{figure}

\subsection{Analysis of Scale-Out and Expert Count}

We evaluate scale-out by deploying 236B DeepSeek-V2 on eight NMP devices and comparing against 8$\times$H100. Because its experts are larger, too few experts fit per stack for effective complementary groups, so we relax the default stack-local constraint and allow each group to span several stacks within one device. Inter-group balance avoids cross-device hotspot collaboration, while physical mapping keeps helper collaboration stack-local whenever possible to use RCL. In Fig.~\ref{fig:scalability}, 3D-MoE achieves throughput speedups of 3.51$\times$ and 1.46$\times$ over the GPU and strongest NMP baselines, respectively.

\begin{sloppypar}
On an 8-expert Llama model~\cite{DavidAU2025Llama3MOEDarkChampion} using NMP-S and 1$\times$H100, 3D-MoE achieves a 1.2$\times$ gain over the best NMP baseline because fewer experts provide less grouping flexibility. This is consistent with our target of modern many-expert models.
\end{sloppypar}

\subsection{Hardware Overhead}
\label{sec:area_overhead}

Table~\ref{table:area_overhead} shows that 3D-MoE adds 2.11\% logic-die area per stack relative to the NMP substrate shared by all schemes~\cite{park2022192,hadidi2017demystifying,yun2024duplex,li2026hardware}. It also consumes 0.337~W of peak power per stack, less than 0.8\% of the 45~W logic-die budget assumed in prior 3D-stacked NMP designs~\cite{pan2025stratum}.

\begin{table}[tbp]
% \vspace{-0.5em}
\caption{Additional area and power of 3D-MoE.}
\vspace{-0.5em}
\centering
\scriptsize
\begin{tabular}{lccc}
\hline
\textbf{Components} & \textbf{Area (mm$^2$)} & \textbf{Area ratio of logic die} & \textbf{Peak power (W)} \\
\hline
DFU   & 0.56 & 0.94\% & 0.172 \\
RCL & 0.156 & 0.28\% & 0.033 \\
DRU & 0.26 & 0.42\% & 0.080 \\
Look-Up Table & 0.28 & 0.47\% & 0.052 \\
\hline
\textbf{Total}  & 1.256 & 2.11\% & 0.337 \\
\hline
\end{tabular}
\label{table:area_overhead}
\vspace{-1em}
\end{table}

\section{Related Work}

\textbf{MoE placement and load balancing.} Existing methods either predict expert demand or react to observed hotspots. MoETuner~\cite{go2025moetuner} uses coarse historical loads; dynamic EP methods~\cite{deepseek_eplb_2025,song2024promoe,zhang2025popfetcher,yao2024exploiting,liu2025netmoe} use replication or remapping with weight-movement overhead; and Yu et al.~\cite{yu2025orders} characterize temporal/spatial selection patterns for data-movement forecasting and GPU/wafer-scale placement. HD-MoE~\cite{huang2025hd} trades EP locality for TP-based hot-expert balancing on 3D NMP, while Stratum~\cite{pan2025stratum} uses topic prediction for tiered placement. The key distinction is the optimization signal: prior methods forecast demand, replicate/relocate experts, or distribute hot experts with TP, whereas 3D-MoE uses offsetting per-layer decode-step loads to place complementary experts in stack-local EP groups, then uses CMN for residual hotspots without domain labels.

\textbf{NMP accelerators for LLM serving.} AttAcc~\cite{park2024attacc}, Duplex~\cite{yun2024duplex}, H$^2$-LLM~\cite{li2025h2}, PAPI~\cite{he2025papi}, and Helios~\cite{li2026hardware} accelerate attention and other bandwidth-bound LLM kernels on NMP. They do not directly address MoE load balancing or stack-local one-to-many assistance for EP hotspots, which are the focus of 3D-MoE.

\section{Conclusion}
This paper presents 3D-MoE, a hardware--software co-designed system for MoE decode on 3D-stacked NMP. To preserve EP locality while mitigating severe, time-varying load imbalance, 3D-MoE combines complementarity-guided grouping and mapping, a lightweight Collaborative Memory-Network for low-cost local hotspot assistance, and online remapping to track workload drift. Across three MoE models, 3D-MoE achieves 1.58$\times$--1.86$\times$ end-to-end throughput improvement over the strongest NMP baseline.

%%%%%%% -- PAPER CONTENT ENDS -- %%%%%%%%

%%%%%%%%% -- BIB STYLE AND FILE -- %%%%%%%%
\bibliographystyle{IEEEtranS}
\bibliography{refs}
%%%%%%%%%%%%%%%%%%%%%%%%%%%%%%%%%%%%

\end{document}